\documentclass[journal]{IEEEtran}
\ifCLASSINFOpdf
\else
\fi
\usepackage{amsmath,graphicx}
\usepackage{icomma}
\usepackage{algorithm}
\usepackage{tabularx}

\usepackage{cite}
\usepackage[ansinew]{inputenc}
\usepackage{psfrag,color}
\usepackage[labelformat=simple]{subcaption}
\usepackage{amsfonts,amssymb,amsthm,amscd,amsmath}
\usepackage{mathtools}
\usepackage{commath,mathabx}
\usepackage{enumerate}
\usepackage{algorithmic}
\usepackage[export]{adjustbox}
\usepackage{textcomp}
\usepackage{xcolor}
\usepackage[overload]{empheq}

\newcommand{\T}{{\scalebox{.65}{$\rm T$}}}

\newcommand{\E}{{\rm E}}
\newcommand{\uM}{{\mathbf u}}

\newcommand{\xM}{{\mathbf x}}

\newcommand{\w}{{\mathbf w}}

\newcommand{\Tr}{{\rm Tr}}

\definecolor{laranja}{rgb}{0.8,0.5,0}
\definecolor{capri}{rgb}{0.0, 0.75, 1.0}
\definecolor{carmine}{rgb}{0.59, 0.0, 0.09}
\definecolor{dimgray}{rgb}{0.41, 0.41, 0.41}

\DeclareMathAlphabet\mathbfcal{OMS}{cmsy}{b}{n}

\newcommand{\boxedeqn}[1]{
	\begin{equation}
	\fbox{$\displaystyle #1 $}
	\end{equation}}

\begin{document}
	
%
\title{A Low-Cost Algorithm for Adaptive Sampling and Censoring in Diffusion Networks}
%
%
%

\author{Daniel G.~Tiglea, Renato Candido, and Magno T.~M.~Silva, \IEEEmembership{Member, IEEE} \thanks{This work was supported by FAPESP under Grant 2017/20378-9, by CNPq under Grants 132586/2018-5 and 304715/2017-4 and by CAPES under Finance Code 001. 
		
Some preliminary parts of this work appeared as conference papers~\cite{Tiglea_Eusipco2019,Tiglea_Eusipco2020}.

The authors are with the Electronic Systems Engineering Department, Escola Politécnica, University of São Paulo, São Paulo, SP, Brazil, e-mails:\{dtiglea,~renatocan,~magno\}@lps.usp.br, ph. +55-11-3091-5134.}}

%
%

\markboth{Submitted to IEEE Transactions on Signal Processing, March~2020}%
{Shell \MakeLowercase{\textit{et al.}}: Bare Demo of IEEEtran.cls for IEEE Journals}
%



\maketitle

\begin{abstract}
Distributed signal processing has attracted widespread attention in the scientific community due to its several advantages over centralized approaches. Recently, graph signal processing has risen to prominence, and adaptive distributed solutions have also been proposed in the area. Both in the classical framework and in graph signal processing, sampling and censoring techniques have been topics of intense research, since the cost associated with measuring and/or transmitting data throughout the entire network may be prohibitive in certain applications. In this paper, we propose a low-cost adaptive mechanism for sampling and censoring over diffusion networks that uses information from more nodes when the error in the network is high and from less nodes otherwise. It presents fast convergence during transient and a significant reduction in computational cost and energy consumption in steady state. As a censoring technique, we show that it is able to noticeably outperform other solutions.  We also present a theoretical analysis to give insights about its operation, and to help the choice of suitable values for its parameters.
	

\end{abstract}


%
\IEEEpeerreviewmaketitle

\section{Introduction} \label{sec:intro}
%
%
%
%
\IEEEPARstart{O}{ver} the last decade, adaptive diffusion networks have become a consolidated tool for distributed parameter estimation and signal processing. Compared to centralized approaches, which require a central unit to receive and process the data from the entire network, this kind of solution presents better scalability, autonomy, and flexibility~\cite{Sayed_Networks2014,lopes2008diffusion,cattivelli2009diffusion,cattivelli2008diffusion,li2009distributed}. 
As a result, adaptive diffusion networks are regarded as effective solutions in a handful of applications, such as target localization and tracking~\cite{Sayed_Networks2014}, spectrum sensing in mobile networks~\cite{di2013bio, Sayed_Networks2014}, medical applications~\cite{akyildiz2002survey}, among others.

These tools consist in a set of connected \emph{agents}, or \emph{nodes}, that are able to collect local data, carry out calculations and communicate with other nearby agents, i.e., its neighbors. The collective goal of the network is to estimate a parameter vector of interest. For this purpose, each node usually computes its own \emph{local} estimate in what is called the \emph{adaptation step}. Then, the neighboring nodes cooperate to reach a \emph{global} estimate of the vector of interest. This stage is usually called the \emph{combination step}. The order in
which the adaptation and combination stages are performed leads to two possible schemes: the adapt-then-combine (ATC) and combine-then-adapt (CTA) strategies. With these two steps, the aim of adaptive diffusion networks is to estimate the parameters of interest without a central processing unit~\cite{Sayed_Networks2014,lopes2008diffusion, cattivelli2009diffusion,cattivelli2008diffusion,li2009distributed,di2013bio,akyildiz2002survey,Fernandez-Bes2017,NassifICASSP2018,HuaEusipco2018}. 

More recently, graph signal processing (GSP) and graph adaptive filtering~\cite{SandryhailaTSP2013,ShumanSPM2013,ChenTSP2015,AnisTSP2016,TsitsveroTSP2016,DiLorenzoTSP2018} have become topics of intense research within the signal processing community, particularly in the field of diffusion networks~\cite{NassifICASSP2018,HuaEusipco2018,DiLorenzoTSP2017}. In comparison with the original distributed adaptation problem, graph adaptive filters incorporate information from the topology of the network in the adaptation step, which is useful in situations where this topology plays an important role in the dynamics of the signals of interest~\cite{NassifICASSP2018,HuaEusipco2018}. This is the case in many network-structured
applications that have emerged in recent years, such as smart grids, internet of things,  transportation and communication networks, among many others
\cite{NassifICASSP2018,HuaEusipco2018,DiLorenzoTSP2017,SandryhailaTSP2013,ShumanSPM2013,ChenTSP2015,AnisTSP2016,TsitsveroTSP2016,DiLorenzoTSP2018}. In these cases, graphs are convenient modeling tools, since they are well suited to represent irregular structures. 


When implementing distributed solutions, it is often desirable to restrict the number of data measurements and the amount of information transmitted across the network. For instance, when these strategies are implemented on wireless sensor networks, where
energy consumption is often the most critical constraint~\cite{takahashi2010link,arroyo2013censoring,fernandez2015censoring}.
Consequently, several solutions have been proposed to reduce the energy consumption associated with the communication between nodes. Some seek to reduce the amount of information sent in each transmission~\cite{arablouei2013distributed,chouvardas2013trading}, whereas others turn links off according to selective communication policies~\cite{lopes2008topologies,takahashi2010link,zhao2012single,xu2015adaptive}. Finally, there are the \emph{censoring} techniques, which seek to avoid the transmission of information from certain nodes to any of their neighbors~\cite{arroyo2013censoring,gharehshiran2013distributed,fernandez2015censoring,yang2018distributed,berberidis2015adaptive}.  Thus, the censored nodes may turn their transmitters off, which saves energy and reduces the amount of information used in the processing~\cite{fernandez2015censoring,berberidis2015adaptive}.

\begin{figure}[htb!]
	\centering
	\includegraphics*[trim=0cm 1.3cm 0cm 0.25cm, clip=true,width=0.8\columnwidth]{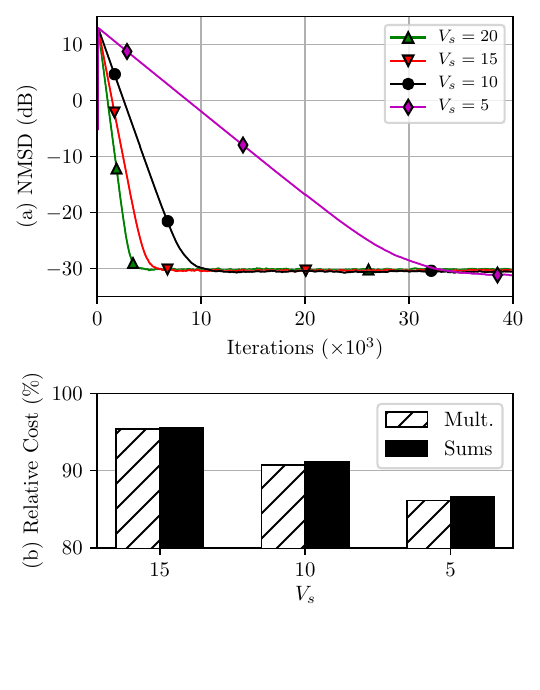}
	\caption{\label{fig:intro} Simulation results obtained for a 20-node network running ATC dNLMS with $V_s$ nodes sampled per iteration. The simulation scenario is described in Section~\ref{sec:simulation}. (a) NMSD curves and (b) Relative number of multiplications and sums in comparison with the case where all $V_s\!=\!20$ nodes are sampled. }
\end{figure}

Furthermore, in certain situations, the cost associated with the measurement and processing of the data  in every node at every time instant is prohibitively high, and thus some sort of \emph{sampling} mechanism is required~\cite{DiLorenzoTSP2018,DiLorenzoTSP2017}. Sampling can greatly reduce the computational cost and memory burden associated with the learning task, but it may also impact the performance of the algorithm. To illustrate this, Fig.~\ref{fig:intro} shows simulation results obtained in a stationary environment considering a network with 20 nodes, which run the ATC diffuse normalized least-mean-square (dNLMS) algorithm~\cite{Sayed_Networks2014,lopes2008diffusion, cattivelli2009diffusion} in conjunction with a sampling technique where
$V_s$ nodes are randomly sampled at every iteration. The results are presented for $V_s \in \{5,\ 10,\ 15,\ 20\}$. The simulation scenario is described in detail in Section~\ref{sec:simulation}, and we adopt the network mean-square-deviation (NMSD) as a performance indicator. To evaluate the computational cost, we present the average number of multiplications and sums per iteration for each value of $V_s$. They are presented as percentages of the number of operations performed when all $V_s\!=\!20$ nodes are sampled. We observe that the less nodes are sampled, the lower the computational cost. Nonetheless, there is a clear impact on the convergence rate, which becomes increasingly slower as the number of sampled nodes decreases. Furthermore, we observe that the steady-state performance is not noticeably affected by the sampling. One intuitive explanation for this is that sampling reduces the rate with which information enters the adaptive network, which leads to a slower convergence rate. However, once the algorithm achieves the steady state, the introduction of more information into the network usually does improve the performance in a stationary environment. 

The question that arises from this experiment is whether it is possible to design a more ``intelligent'' sampling strategy, in which more nodes are sampled when the estimation error is high (e.g., during transient) and less nodes otherwise, thus preserving the convergence rate of the algorithm. In this paper, we propose such a technique. It can greatly reduce the computational cost during steady state while maintaining transient performance. Moreover, with slight modifications it can also be employed as a censoring strategy, allowing the nodes to save energy by transmitting less information to their neighbors. In particular, we show that the censoring version of the proposed technique is able to outperform other state-of-the-art censoring mechanisms~\mbox{\cite{arroyo2013censoring,gharehshiran2013distributed}}.

The paper is organized as follows. The general formulation of diffuse adaptive networks is presented in Section~\ref{sec:problem} for both the classical distributed estimation problem and for GSP. In Section~\ref{sec:sampling}, the adaptive sampling mechanism is introduced, and we analyze its behavior in Section~\ref{sec:analysis}. In Section~\ref{sec:cost}, the computational cost reduction of the proposed sampling mechanism is analyzed in more detail. Finally, simulation results are presented in Section~\ref{sec:simulation}, and Section~\ref{sec:conclusoes} closes the paper with the main conclusions and ideas for future work.


\noindent \textit{Notation}. We use normal font letters to denote scalars, boldface lowercase letters for vectors, and boldface uppercase letters for matrices. Moreover, $[\xM]_k$ denotes the $k$-th entry of the vector $\xM$, and if $\mathcal{X}$ is a set, $|\mathcal{X}|$ denotes its cardinality. Finally, $(\cdot)^{\T}$ denotes transposition, $\E\{\cdot \}$ the mathematical expectation, $\Tr[\cdot]$ the trace of a matrix, and $\|\cdot\|$ the Euclidean norm.

\section{Problem Formulation} \label{sec:problem}

\begin{figure}[htb!]
	\centering
	\includegraphics*[width=1\columnwidth]{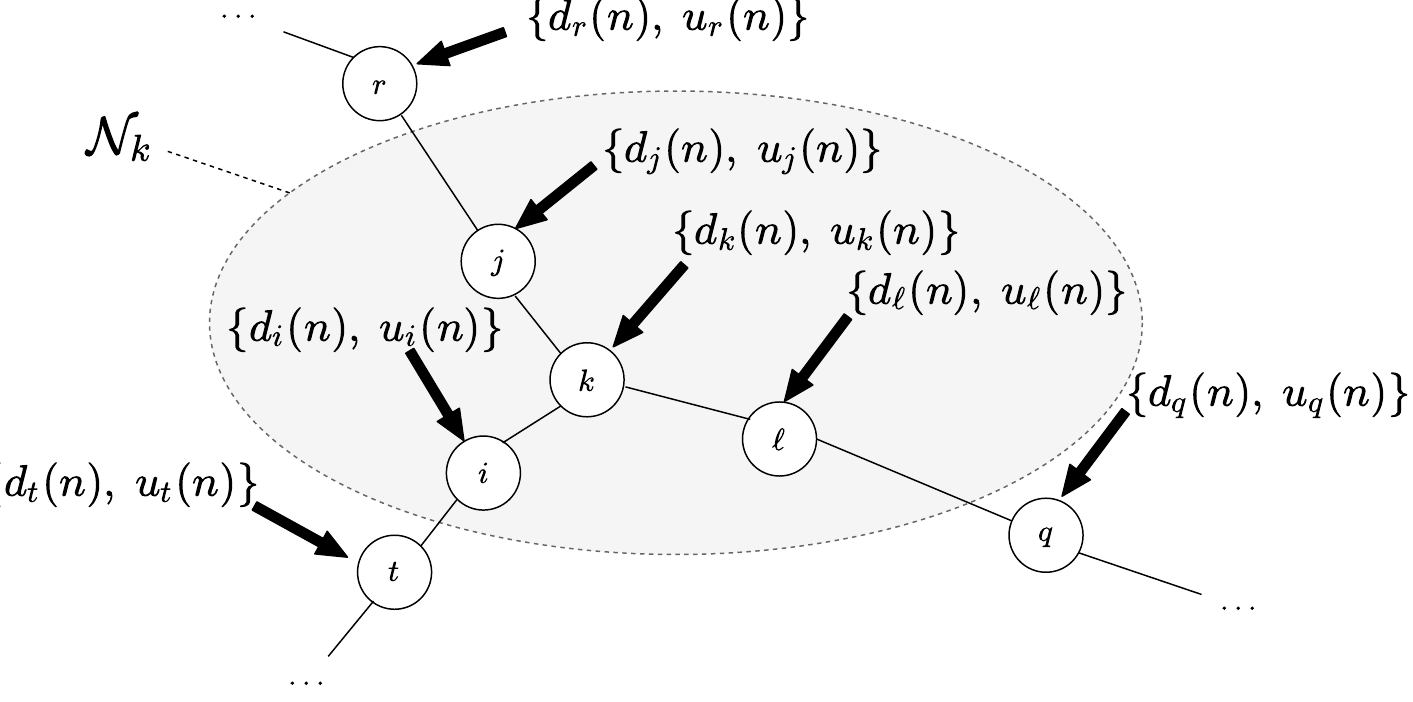}
	\caption{\label{fig:network1} Example of a diffusion network. In this case, the neighborhood of node $k$ consists of the nodes $i$, $j$, $k$, and $\ell$.}
\end{figure}

Let us consider a network with a predefined topology and $V$ nodes labeled $1,\ \cdots,\ k,\ \cdots,\ V$. Two nodes are considered neighbors if they can exchange information, and we denote by $\mathcal{N}_k$ the neighborhood of node $k$ including $k$ itself.
Furthermore, as depicted in Fig.~\ref{fig:network1}, each node $k$ has access to an input signal $u_k(n)$ and to a desired signal $d_k(n)$, given by~\cite{Sayed_Networks2014,lopes2008diffusion, cattivelli2009diffusion,NassifICASSP2018,HuaEusipco2018}
\begin{equation} \label{eq:dk}
d_k(n) = \xM_k^{\T}(n)\mathbf{w}^{\rm o} + v_k(n), 
\end{equation}
where $v_k(n)$ is the measurement noise at node $k$, which is assumed to be independent of the other variables and zero-mean with variance $\sigma^2_{v_k}$, and  $\mathbf{w}^{\mathrm{o}}$ and $\xM_k(n)$ are $M$-length column vectors that represent respectively the optimal system and a processed version of the input signal $u_k(n)$. 

\begin{figure}[htb!]
	\centering
	\includegraphics*[width=0.9\columnwidth]{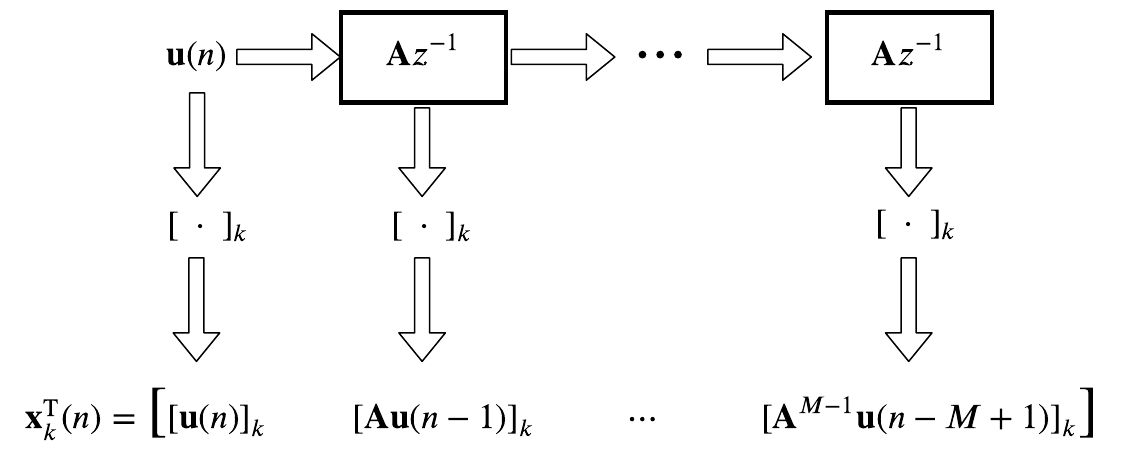}
	\caption{\label{fig:x} Obtaining $\xM_k(n)$ from $\uM(n)$ for each node $k$.}
\end{figure}

In the classical adaptation problem, $\xM_k(n)$ is usually considered to be a regressor vector, given by~\cite{Sayed_Networks2014,lopes2008diffusion, cattivelli2009diffusion} 
\begin{equation} \label{eq:classic}
	\xM_k(n) = [u_k(n)\,  \ u_k(n\!-\!1)\, \cdots\, \ u_k(n\!-\!M\!+\!1)]^{\T}.
\end{equation}
Thus, $d_k(n)$ can be seen as a noisy measurement of the output of the finite impulse response (FIR) optimal filter $\mathbf{w}^\mathrm{o}$. In the context of graph adaptive filtering, $\xM_k(n)$ is assumed to be related to the topology of the graph through~\cite{NassifICASSP2018,HuaEusipco2018}
\begin{equation} \label{eq:graph}
	\!\!\xM_k(n)\! \!=\!\! \left[[\mathbf{u}(n)]_{k}\ [\!\mathbf{A}\hspace*{-0.04cm}\mathbf{u}(n\!\!-\!\!1)]_k\ \cdots\ [\!\mathbf{A}^{\!\!M\!-\!1}\hspace*{-0.04cm}\mathbf{u}(n\!\!-\!\!M\!\!+\!\!1)]_k\hspace*{-0.04cm}\right]^{\! \rm T}\hspace*{-0.04cm},
\end{equation}
where $\mathbf{u}(n) = [u_1(n) \ \,  u_2(n) \, \ \cdots \, \ u_V(n)]^{\T}$, and $\mathbf{A}$ is the graph shift operator. Possible choices for $\mathbf{A}$ include the adjacency matrix, the graph Laplacian matrix, among others~\mbox{\cite{NassifICASSP2018,HuaEusipco2018}}. The relation between $\xM_k(n)$ and $\uM(n)$ in~\eqref{eq:graph} is illustrated in Fig.~\ref{fig:x} and can be interpreted as follows: $\mathbf{u}(n)$ represents the ``raw'' information available at each node of the network at the iteration $n$, whereas $\xM_k(n)$ models the spreading of that information throughout the graph, which is the result of both a temporal and spacial shift, or ``delay''. Moreover, $\mathbf{w}^{\rm o}$ models how exactly the graph topology and time lag affect the spreading of information, and $d_k(n)$ represents a noisy measurement of the information available at node $k$ as a result of this spreading process~\cite{NassifICASSP2018,HuaEusipco2018}. We should notice that there is a clear analogy to the tapped delay line commonly found in discrete-time filters~\cite{SayedL2008}.

The difference between the classic framework and the graph-based one lies in the role of the spatial aspect of the problem.  In the former, the topology of the network does not influence the dynamics of the desired signal. Thus, $d_k(n)$ depends only on the signal $u_k(n)$ and on the measurement noise $v_k(n)$, and is independent of $u_i(n)$ for all $i \!=\! 1,\cdots,V,\ i \!\neq\! k$. This occurs since the information does not ``travel'' from one node to another. In graph signal processing, if the nodes $i$ and $k$ are immediate neighbors, $d_k(n)$ does depend on $u_i(n\!-\!1)$, since the information from one node spreads to its neighbors over time. Moreover, if nodes $j$ and $k$ are two-hop neighbors (i.e., it is possible to travel from node $j$ to node $k$ in two hops), $d_k(n)$ also depends on $u_j(n\!-\!2)$, and so forth. Hence, the topology of the network plays a major role in how the desired signal $d_k(n)$ unfolds at each node $k$. This makes graph adaptive filtering well suited for distributed problems where both time and space must be taken into consideration, e.g., meteorology~\cite{NassifICASSP2018,HuaEusipco2018}.
Nonetheless, despite the conceptual differences between both applications, in all cases Model~\eqref{eq:dk} is assumed to hold. Thus a common mathematical formulation can be used to describe them to a certain extent.

In both situations, the objective of the network is to obtain an estimate $\w$ of $\mathbf{w}^\mathrm{o}$ in a distributed manner by solving~\mbox{\cite{Sayed_Networks2014,lopes2008diffusion, cattivelli2009diffusion,NassifICASSP2018,HuaEusipco2018}}
\begin{equation} \label{eq:minimize}
\min_{\mathbf{w}} J(\w)\!=\!\min_{\mathbf{w}}\textstyle\sum_{k=1}^{V}J_k(\w),
\end{equation}
where $J_k(\w)$ are the local costs at each node $k$, given by
\begin{equation} \label{eq:cost_w}
J_k(\w)\!\triangleq\!\E\{|d_k(n)\!-\!\xM_k^{\T}(n)\w|^2\}.
\end{equation}
Thus, at each iteration, every node $k$ calculates a local estimate of $\w^{\rm o}$ in order to minimize its individual cost function $J_k(\w)$. This is done by using only the data available locally, as well as the information transmitted by neighboring nodes. Then, the nodes cooperate to form the global estimate $\w$. It can be shown that, when the combination of the local estimates is done properly, they converge to a single common solution~\mbox{\cite{Sayed_Networks2014,lopes2008diffusion,cattivelli2008diffusion, cattivelli2009diffusion}.}

Several adaptive solutions have been proposed in the literature to solve~\eqref{eq:minimize}, one of them being the ATC dNLMS algorithm~\cite{Sayed_Networks2014,lopes2008diffusion, cattivelli2009diffusion, NassifICASSP2018,HuaEusipco2018}. The adaptation and combination steps of this algorithm are respectively given by
\begin{subequations} \label{eq:lms_sayed}
	\begin{empheq}[left={\empheqlbrace\,}]{align}
	&\boldsymbol{\psi}_k(n+1)\!=\!\w_k(n)\!+\!\mu_k(n) \xM_k(n) e_k(n) \label{eq:lms_sayed1}\\
	&\w_k(n+1)\!=\!\textstyle\sum_{j \in \mathcal{N}_k} c_{jk} \boldsymbol{\psi}_j(n+1), \label{eq:lms_sayed2}
	\end{empheq}
\end{subequations}
where $\boldsymbol{\psi}_k$ and  $\w_k$ are the local and combined estimates of $\w^{\rm o}$ at node $k$,
\begin{equation} \label{eq:ek}
e_k(n) = d_k(n) - \xM_k^{\T}(n) \w_k(n)
\end{equation}
is the estimation error, and
\begin{equation} \label{eq:muk}
\mu_k(n) \! =\! \frac{\widetilde{\mu}_k}{\delta + \|\xM_k(n)\|^2}
\end{equation}
is a normalized step size with $0\!<\!\widetilde{\mu}_k\!<\!2$ and a
small regularization factor $\delta\!>\!0$~\cite{Sayed_Networks2014}. Moreover, $\{c_{jk}\}$ are combination weights satisfying~\cite{lopes2008diffusion, cattivelli2009diffusion}
\begin{equation} \label{eq:ck}
c_{jk}\!\geq\!0,\ \textstyle\sum_{j\in \mathcal{N}_k}\!c_{jk}\!=\!1,\ \text{and}\ c_{jk}\!=\!0 \ \text{for}\ j\notin \mathcal{N}_k.
\end{equation}
Possible choices for $\{c_{jk}\}$ include the Uniform, Laplacian,
Metropolis, and Relative Degree rulese~\cite{Sayed_Networks2014,cattivelli2008diffusion}, as well as adaptive
schemes~\cite{Fernandez-Bes2017,takahashi2010diffusion,yu2013strategy}, such as the Adaptive Combination Weights (ACW) algorithm~\cite{takahashi2010diffusion,tu2011optimal}. ACW incorporates information from the noise profile across the network, and is obtained by solving an optimization problem with respect to $\{c_{jk}\}$~\cite{takahashi2010diffusion,tu2011optimal}. Its equations are given by~\cite{tu2011optimal}
\begin{equation} \label{eq:acw1}
c_{jk}(n)= 
\begin{cases}
\frac{\widehat{\sigma}^{-2}_{jk}(n)}{\sum_{\ell \in \mathcal{N}_k}\widehat{\sigma}^{-2}_{\ell k}(n)}\ \text{if}\ j \in \mathcal{N}_k \\
0, \ \text{otherwise}
\end{cases},
\end{equation}
where $\sigma^{2}_{jk}$ is updated as
\begin{equation} \label{eq:acw2}
\widehat{\sigma}^2_{jk}(n)\! =\! (1\!-\!\nu_k)\widehat{\sigma}^2_{jk}(n\!-\!1)\!+\!\nu_k \lVert\boldsymbol{\psi}_j(n\!+\!1)\!-\!\w_k(n)\rVert^2,
\end{equation}
with $\nu_k\!>\!0$ for $k\!=\!1,\cdots,V$. Hence, greater weights are assigned to the nodes with smaller noise variances~\cite{tu2011optimal}. We should notice that $\{c_{ik}(n)\}$ defined by~\eqref{eq:acw1} satisfy~\eqref{eq:ck}. To avoid division by zero, in this paper we adopt a regularized version of~\eqref{eq:acw1}, i.e., we replace $\widehat{\sigma}^{-2}_{jk}(n)$ and $\widehat{\sigma}^{-2}_{\ell k}(n)$ by $[\delta_c+\widehat{\sigma}^2_{jk}(n)]^{-1}$ and $[\delta_c+\widehat{\sigma}^{2}_{\ell k}(n)]^{-1}$ in~\eqref{eq:acw1}, respectively, where $\delta_c\!>\!0$ is a small constant.


Finally, it is worth recalling that we could also employ a CTA strategy~\cite{Sayed_Networks2014,lopes2008diffusion, cattivelli2009diffusion,cattivelli2008diffusion,li2009distributed,di2013bio,akyildiz2002survey,Fernandez-Bes2017,NassifICASSP2018,HuaEusipco2018} in conjunction with other adaptive solutions~\cite{cattivelli2008diffusion,li2009distributed}. For simplicity, in this paper we will only consider the ATC strategy with the dNLMS algorithm. However, the results can be straightforwardly extended to other approaches.

\section{The sampling algorithm}\label{sec:sampling}
At each iteration, the ATC dNLMS algorithm estimates the vector
$\w^{\rm o}$ from the data $\{d_k(n), u_k(n)\}$. In our sampling proposal, we
define the variable $\widebar{s}_k(n)$ that assumes the values zero or one to decide
if each node $k$ should be sampled and if~\eqref{eq:lms_sayed1} should be computed or not. Thus, we
recast~\eqref{eq:lms_sayed1}~as
\boxedeqn{ \label{eq:lms_sayed_mod}
	\boldsymbol{\psi}_k(n+1) = \w_k(n) + \widebar{s}_k(n) \mu_k(n)  \xM_k(n) e_k(n).
}
If $\widebar{s}_k(n)\!=\!1$, $d_k(n)$ is sampled, $e_k(n)$ is computed as in \eqref{eq:ek} and \eqref{eq:lms_sayed_mod} coincides with \eqref{eq:lms_sayed1}. In contrast, if $\widebar{s}_k(n)\!=\!0$, $d_k(n)$ is not sampled, $\xM_k^{\T}(n) \w_k(n)$, $e_k(n)$ and $\mu_k(n)$ are not computed, and \mbox{$\boldsymbol{\psi}_k(n\!+\!1)\!=\!\w_k(n)$.} 

To determine $\widebar{s}_k(n)$, we define $s_k(n) \! \in \! [0,1]$ such that 
\begin{equation} \label{eq:s_barra_s}
\widebar{s}_k(n) = \begin{cases}
1,\ \text{if} \ s_k(n) \geq 0.5, \\
0, \ \text{otherwise}
\end{cases}.
\end{equation}
We then minimize the following cost function with respect to $s_k(n)$:
\begin{equation} \label{eq:cost}
J_{s,k}(n) = [s_k(n)]\beta \bar{s}_k(n)\!+\!\left[1\!-\!s_k(n)\right] \!\!\sum_{j \in \mathcal{N}_k}\!\! c_{ik}(n)e_i^2(n),
\end{equation}
where  $\beta \!\! > \!\! 0$ is a parameter introduced to control how much the sampling of the nodes is penalized.
When the error is high in magnitude or when node $k$ is not being sampled ($\bar{s}_k\!=\!0$), $J_{s,k}(n)$ is minimized by making $s_k(n)$ closer to one, leading to the sampling of node $k$. This ensures that the algorithm keeps sampling the nodes while the error is high and resumes the sampling of
idle nodes at some point, enabling it to detect changes in the environment. In contrast, when node $k$ is being sampled (\mbox{$\bar{s}_k\!=\!1$}) and the error is small in magnitude in comparison to $\beta$, $J_{s,k}(n)$ is
minimized by making $s_k(n)$ closer to zero, which leads the algorithm to stop sampling node $k$. This desirable behavior depends on a proper choice for $\beta$, which is addressed in Section~\ref{sec:analysis}.

Inspired by convex combination of adaptive filters (see~\cite{SPM2016,Garcia_Biased2010} and their references),
rather than directly adjusting $s_k(n)$, we update an auxiliary
variable $\alpha_k(n)$ related to it via \cite{Garcia_Biased2010}
\begin{equation} \label{eq:tb}
s_k(n) = \phi[\alpha_k(n)] \triangleq \frac{\mathrm{sgm}[\alpha_k(n)]-\mathrm{sgm}[-\alpha^+]}{\mathrm{sgm}[\alpha^+]-\mathrm{sgm}[-\alpha^+]},
\end{equation}
where $\mathrm{sgm}[x]\!=\!(1\!+\!e^{-x})^{-1}$
is a sigmoidal function and $\alpha^{\! +}$ is the maximum value $\alpha_k$ can assume.
We should notice that $\phi[\alpha^{\! +}\!] \! = \! 1$, $\phi[0] \!=\! 0.5$, and $\phi[-\alpha^{\! +}\!]\!=\!0$. In the literature, $\alpha^{\! +}\!=\!4$ is usually adopted~\cite{Garcia_Biased2010}. 

By taking the derivative of \eqref{eq:cost} with respect to $\alpha_k(n)$, we obtain
the following stochastic gradient descendent rule:
\begin{equation} \label{eq:b_original}
\begin{aligned}
\alpha_k(n\!+\!1)&\!=\!\alpha_k(n)+\\
&\mu_s\phi^{\prime}[\alpha_k(n)]\!\left[\!\sum_{i \in \mathcal{N}_k}\!\!c_{ik}(n) e_i^2(n)\!-\!\beta \bar{s}_k(n) \right],
\end{aligned}
\end{equation}
where $\mu_s>0$ is a step size and
\begin{equation} \label{eq:phi_linha}
\phi^{\prime}[\alpha_k(n)]\!\triangleq\!\frac{d s_k(n) }{d\alpha_k(n)}\!=\! \frac{\mathrm{sgm}[\alpha_k(n)]\{1\!-\!\mathrm{sgm}[\alpha_k(n)]\}}{\mathrm{sgm}[\alpha^+]\!-\!\mathrm{sgm}[-\alpha^+]}.
\end{equation}

Equation~\eqref{eq:b_original} cannot be used for sampling since it requires the errors to be computed  to decide if the nodes should be sampled or not,
which is contradictory. To address this issue, we replace $e_i(n)$ in~\eqref{eq:b_original} by its latest measurement we have access to,
which is denoted by $\varepsilon_i(n)$. When the node is sampled, $\varepsilon_i(n)\!=\!e_i(n)$. We thus obtain 
\boxedeqn{ \label{eq:b_modificada}
	\begin{aligned}
	\alpha_k(n\!+\!1)&\!=\!\alpha_k(n)+ \\
	&\mu_s\phi^{\prime}[\alpha_k(n)]\! \left[\!\sum_{i \in \mathcal{N}_k}\!\!c_{ik}(n) \varepsilon_i^2(n)\!-\!\beta  \widebar{s}_k(n)  \!\right].
	\end{aligned}
}

Equation~\eqref{eq:b_modificada} is the foundation of the adaptive sampling mechanism. In conjunction with~\eqref{eq:lms_sayed_mod}, it leads to an adaptive-sampling version of the dNLMS algorithm, named as adaptive-sampling diffusion NLMS (AS-dNLMS). This algorithm is summarized in Table~\ref{tab:asdnlms}. Since~\eqref{eq:b_modificada} depends only on the estimation error at each sampled node, the proposed sampling technique can be extended to any adaptive diffuse algorithm.

It is interesting to notice that although we used $s_k(n)$ in the derivation of the algorithm, it does not have to be calculated explicitly, since it does not arise in~\eqref{eq:lms_sayed_mod} or~\eqref{eq:b_modificada}. Instead, only $\widebar{s}_k(n)$ and $\frac{ds_k(n)}{d\alpha_k(n)}$ appear. The latter can be stored in a look-up table, and the former  is related to $\alpha_k(n)$ by
\begin{equation} \label{eq:s_barra_alpha}
\widebar{s}_k(n) = \begin{cases}
1,\ \text{if} \ \alpha_k(n) \geq 0, \\
0, \ \text{otherwise}
\end{cases},
\end{equation}
as can be seen from~\eqref{eq:s_barra_s} and~\eqref{eq:tb}. 

\begin{table}[h!tb]
	\centering
	\caption{Summary of the AS-dNLMS algorithm} \label{tab:asdnlms}
	\begin{tabularx}{\columnwidth}{|X|}
		
		\hline
		
		\textit{\% Initialization} 

		 \hspace*{0.5cm} For each node $i \!=\! 1, \cdots,V$, set $\alpha_i(0)\! \gets \! \alpha^+, \widebar{s}_i(0)\!\gets\!1$, $\varepsilon_i(n)\!\gets\!0,$\\
		 
		 \hspace*{0.5cm} $\xM_i(n) = \mathbf{0}$, $\boldsymbol{\psi}_i(0) \!\gets\! \mathbf{0}$, $\w_i(0) \!\gets\! \mathbf{0}$. 
		 
		 \textit{\% Then, repeat the following for every iteration $n\! \geq\! 0$ and every node $k$}: 
		
		\textit{\% Adaptation Step} 

		\hspace*{0.5cm} \textbf{If} $\alpha_k(n)\!\geq\!0$, \textbf{do} : \\

		\hspace*{0.5cm} \hspace*{0.5cm} $ \widebar{s}_k(n) \! \gets \! 1$ \\
		
		\hspace*{0.5cm} \textbf{Else, do}: \\
		
		\hspace*{0.5cm} \hspace*{0.5cm} $\widebar{s}_k(n) \gets 0$ \\
		
		\hspace*{0.5cm} \textbf{End}
		
		\hspace*{0.5cm} \textbf{If} $\widebar{s}_k(n)\!=\!1$, \textbf{do} : \\
		
		\hspace*{0.5cm} \hspace*{0.5cm} Update $\xM_k(n)$ and $\norm{\xM_k(n)}^2$ \\
		
		\hspace*{0.5cm} \hspace*{0.5cm}  $\mu_k(n) \gets {\widetilde{\mu}_k}/{[\delta + \|\xM_k(n)\|^2]}$
		
		\hspace*{0.5cm} \hspace*{0.5cm} $e_k(n) \gets d_k(n) - \xM_k^{\T}(n) \w_k(n)$ \\
		
		\hspace*{0.5cm} \hspace*{0.5cm} $\varepsilon_k(n)\!\gets\!e_k(n)$ \\
		
		\hspace*{0.5cm} \hspace*{0.5cm} $\boldsymbol{\psi}_k(n+1) \gets \w_k(n) + \widebar{s}_k(n) \mu_k(n)  \xM_k(n) e_k(n)$ \\
		
%
%
		\hspace*{0.5cm} \textbf{Else, do}: \\
		
		\hspace*{0.5cm} \hspace*{0.5cm} $\boldsymbol{\psi}_k(n+1) \gets \w_k(n)$ \\
		
		\hspace*{0.5cm} \textbf{End}

		\textit{\% Transmission} 
		
		\hspace*{0.5cm} Transmit $\boldsymbol{\psi}_k(n+1)$ and $\varepsilon_k^2(n)$ to every node $ \in \mathcal{N}_k$ 
		
		\textit{\% Combination Step} 
		
		
		\hspace*{0,5cm} $\alpha_k(n\!+\!1)\!\gets\!\alpha_k(n)\!+\!\mu_s\phi^{\prime}[\alpha_k(n)] \left[\sum_{i \in \mathcal{N}_k} c_{ik}(n) \varepsilon_i^2(n)\!-\!\beta  \widebar{s}_k(n)  \right]$\vspace*{0.1cm}\\ 

\hspace*{0.5cm} $\w_k(n+1)\!\gets\!\sum_{j \in \mathcal{N}_k} c_{jk}(n) \boldsymbol{\psi}_j(n+1)$ \\
		
		\hline
		
	\end{tabularx}
	
\end{table}

The proposed mechanism reduces the number of sampled nodes in steady state,  decreasing the computational cost. If $\beta$ is chosen appropriately, this  reduction does not occur in the transient and the adaptive-sampling version of the algorithm maintains the same
convergence rate as that of the original with no sampling mechanism. This comes at the expense of a
slight increase of the cost during the transient, since the sampling algorithm requires the computation of an additional update equation per node per iteration. This will be explored in more detail in Section~\ref{sec:cost}. Furthermore, we should mention that when the node $i$ is sampled, it is required to transmit $\varepsilon_i^2(n)=e_i^2(n)$ to its neighbors. Nonetheless, this information can be
sent bundled with the local estimates
$\boldsymbol{\psi}_i$ so as to not increase the number of transmissions.

Finally, we remark that the algorithm described in Table~\ref{tab:asdnlms} can be implemented in conjunction with any rule for the selection of combination weights.  If an adaptive scheme for such selection  is employed, the update of $\{c_{ik}(n)\}$ should  also be included in Table~\ref{tab:asdnlms}. Particularly, if  ACW is considered in conjunction with AS-dNLMS and the sampling of node $k$ ceased for a long period of time, the sampling mechanism could potentially harm the update of the combination weights. This occurs since in this case $\widehat{\sigma}_{kk}$ could tend towards zero in~\eqref{eq:acw2} due to $s_k$ being equal to zero in~\eqref{eq:lms_sayed_mod}. To avoid this, for $j\!=\!k$, we replace $\boldsymbol{\psi}_j(n+1)$ in~\eqref{eq:acw2} by $\widebar{\boldsymbol{\psi}}_k(n+1) \triangleq \widebar{s}_k(n) \boldsymbol{\psi}_k(n+1)+ [1-\widebar{s}_k(n)]\widebar{\boldsymbol{\psi}}_k(n)$.

\subsection*{The Adaptive Sampling Algorithm as a Censoring Strategy} \label{sec:transmi}

With a very simple modification, the proposed adaptive sampling mechanism can also be used as a censoring strategy. This alternate version of AS-dNLMS is obtained by not updating $\boldsymbol{\psi}_k$ at all when node $k$ is not sampled. In other words, instead of using~\eqref{eq:lms_sayed_mod}, we apply
\boxedeqn{ \label{eq:b_censor}
	\begin{aligned}
	\boldsymbol{\psi}_k(n+1) &= [1\!-\!\widebar{s}_k(n)]\boldsymbol{\psi}_k(n) + \\
	&\widebar{s}_k(n) \left[\w_k(n) \!+\!  \mu_k(n)  \xM_k(n) e_k(n)\right].
	\end{aligned}
}


Assuming that the nodes can store past information from their neighbors,
this allows us to cut the number of communications between nodes, since in this case $\boldsymbol{\psi}_k$ and $\varepsilon^2_k$
remain static when $\bar{s}_k\!=\!0$ and there is no need for node $k$ to retransmit them. Thus, when node $k$ is not
sampled in this version of the algorithm, it only receives data and carries out~\eqref{eq:lms_sayed2}, and can therefore turn its transmitter off. This version of the proposed algorithm is named as adaptive-sampling-and-censoring diffusion NLMS (ASC-dNLMS), and it features a lower energy consumption as well as a  computational~cost reduction in comparison with the original dNLMS algorithm.

\section{Theoretical Analysis} \label{sec:analysis}

In the current section, we conduct a theoretical analysis of the proposed sampling mechanism. In particular, we  study the effects of the parameters $\beta$ and $\mu_s$ on its behavior and obtain rules to help select them in a suitable manner. In~\ref{sec:beta}, we show how to choose $\beta$ so as to ensure that the nodes cease to be sampled at some point during steady state. Then, in~\ref{sec:number} we study in more detail how its choice influences the expected number of sampled nodes per iteration. Finally, in~\ref{sec:mus}, we analyze how fast the nodes cease to be sampled depending on the choice for $\mu_s$, and how to select this parameter appropriately based on that information.

\subsection{The parameter $\beta$ and its effects on the algorithm} \label{sec:beta}

The parameter $\beta$ plays a crucial role in the behavior of the AS-dNLMS. It influences the expected number of sampled nodes during steady state, and determines when the sampling mechanism begins to act. Thus, in this section we study its effects on the algorithm and analyze how to select it properly.

Firstly, we study how to choose $\beta$ so that we can ensure that every node will cease to be sampled at some point during steady state. To do so, we examine~\eqref{eq:b_modificada} while node $k$ is being sampled. In this case, $\varepsilon^2_i(n)$ and $\beta \bar{s}_k(n)$ can be replaced by $e^2_i(n)$ and $\beta$, respectively. Then, subtracting $\alpha_k(n)$ from both sides in~\eqref{eq:b_modificada} and taking expectations, we get
\begin{equation} \label{eq:expected_b}
\!\!\E\{\Delta \alpha_k(n) \}\!=\!\mu_s  \E \left\{\!\! \phi^{\prime}[\alpha_k(n)]\! \left[ \!\sum_{i \in \mathcal{N}_k}\!\!c_{ik}(n)e_i^2(n)\!-\!\beta \right]\!\right\}\!.
\end{equation}
where  $\Delta \alpha_k(n) \triangleq \alpha_k(n\!+\!1)-\alpha_k(n)$.
To make the analysis more tractable,
$\phi^{\prime}[\alpha_k(n)]$ and the term between brackets in~\eqref{eq:expected_b} are  assumed to be statistically independent.
Although this assumption may seem unrealistic, simulation results suggest it is a reasonable approximation.
Thus, we can write
\begin{equation} \label{eq:expected_b2}
\begin{aligned}
\E\{\Delta \alpha_k(n) \}=&\mu_s\E\{\phi^{\prime}[\alpha_k(n)]\} \times \\
		&\left[\!\sum_{i \in \mathcal{N}_k}\!c_{ik}(n) \E\{e_i^2(n)\}-\beta\right].
\end{aligned}
\end{equation}

In order to stop sampling node $k$,
$\alpha_k(n)$ should decrease along the iterations until it becomes negative.
Since $\phi^{\prime}[\alpha_k(n)]$ is always positive, to
enforce $\E\{\Delta \alpha_k(n)\}$ to be negative while node $k$ is sampled, $\beta$ must
satisfy 
\begin{equation} \label{eq:beta1}
\beta > \textstyle\sum_{i \in \mathcal{N}_k}\!\!c_{ik}(n) \E\{e_i^2(n)\}.
\end{equation}

Assuming that the order of the adaptive filter is sufficient and that the $\tilde{\mu}_k$, $k\!=\!1,\ \cdots,\ V,$ are chosen properly so that the gradient noise can be disregarded, it is reasonable to assume that, during steady state, $\E\{e_i^2(n)\} \approx \sigma_{v_{i}}^2$, which leads to

\begin{equation}
\label{eq:menorque}
\sigma^2_{\min} \leq \textstyle\sum_{i \in \mathcal{N}_k}\!\!c_{ik}(n) \E \{e_i^2(n) \} \leq \sigma^2_{\max},
\end{equation}
where $\sigma^2_{\min}\triangleq \min_{i} \sigma_{v_{i}}^2$, and $\sigma^2_{\max} \triangleq \max_{i} \sigma_{v_{i}}^2$, $i\!=\!1,\cdots,V$.
Thus, the condition
\boxedeqn{ \label{eq:betamin}
	\beta > \sigma^2_{\min}
}
is necessary (but not sufficient) if we wish to stop sampling the nodes at some point during steady state. On the other hand, 
\boxedeqn{ \label{eq:beta2}
	\beta > \sigma^2_{\max}
}
is a sufficient (although not necessary) condition to ensure a reduction in the number of sampled nodes. Moreover, this ensures that \emph{every node} will cease to be sampled at some iteration during steady state in the mean. When $\sigma_{\min}^2=\sigma_{\max}^2$, i.e., every node is subject to the same level of noise power,~\eqref{eq:betamin} and~\eqref{eq:beta2} coincide and form a necessary and sufficient condition.

Moreover, given a certain value of $\beta$, we can analyze when the sampling mechanism will begin to act in terms of the mean-squared error (MSE). From~\eqref{eq:expected_b2} we observe that $\E\{\Delta \alpha_k(n)\} \!\geq\! 0$ as long as $\sum_{i \in \mathcal{N}_k}\!\!c_{ik}(n) \E\{e_i^2(n)\} \!\geq\! \beta$. Since we do not allow $\alpha_k(n)$ to become greater than $\alpha^+$, we conclude that $\E\{\alpha_k(n)\} \!=\! \alpha^+$ for $k\!=\!1,\cdots,V$ as long as $\text{MSE}_{\min}(n) > \beta$, where $\text{MSE}_{\min}(n) \triangleq \min_i \E\{e_i^2(n)\}$, $i \!=\! 1, \cdots, V$. Thus, in the mean, the sampling mechanism does not act as long as the lowest mean-square error in the network remains greater than $\beta$. Consequently, no node will cease to be sampled in the mean during that period. Moreover, the greater the $\beta$, the sooner $\E\{\alpha_k(n)\}$ begins to decrease for $k\!=\!1,\cdots,V$, and the sooner the nodes cease to be sampled.

\subsection{The expected number of sampled nodes} \label{sec:number}

Based on the previous section, we can estimate upper and lower bounds for the expected number $V_s$ of sampled nodes in steady state. For this purpose, we consider each $\widebar{s}_k(n)$ as an independent Bernoulli random variable during steady state that is equal to one with probability $p_{s_k}$ or to zero with probability $1\!-\!p_{s_k}$ for $k\!=\!1, \cdots,V$, with $0 \!\leq\! p_{s_k}\! \leq\! 1$. Thus,
\begin{equation} \label{eq:bounds}
V p_{s_{\min}} \leq  \E\{V_s\} \leq V p_{s_{\max}},
\end{equation}
where $p_{s_{\min}}$ and $p_{s_{\max}}$ are upper and lower bounds for $p_{s_k}$, $k=1,\cdots,V$.

It is useful to note that the sampling mechanism exhibits a cyclic behavior in steady state. Hence, we could approximate $p_{s_k}$ by the expected ``duty cycle'' of the mechanism, i.e.,
\begin{equation} \label{eq:psk}
\displaystyle \widehat{p}_{s_k} = \frac{\theta_k}{\theta_k+\overline{\theta}_k},
\end{equation}
where $\theta_k$ denotes the expected number of iterations per cycle in which node $k$ is sampled and $\overline{\theta}_k$ is the expected number of iterations in which it is not.
Since we are only interested in estimating $p_{s_{\min}}$ and $p_{s_{\max}}$, we do not need to evaluate~\eqref{eq:psk} for every $k$.
Instead, we only need to estimate upper and lower bounds for $\theta_k$~and~$\overline{\theta}_k$, which we respectively denote by $\theta_{\max}$, $\theta_{\min}$, $\overline{\theta}_{\max}$ and $\overline{\theta}_{\min}$. 

For the sake of brevity, in this section we omit the intermediate calculations and skip to the final results  concerning the estimation of these parameters. Nonetheless, a complete demonstration is provided in Appendix~\ref{sec:demonstracao}.

Assuming that we can write
\begin{equation}
\label{eq:menorque2}
\sigma^2_{\min} \leq \textstyle\sum_{i \in \mathcal{N}_k}\!\!c_{ik}(n) \E \{\varepsilon_i^2(n) \} \leq \sigma^2_{\max}
\end{equation}
for $k=1,\cdots,V$ during steady state, we can estimate $\theta_{\max}$ by finding the maximum number of iterations any node can remain sampled in the mean. Considering a worst-case scenario, as well as the fact that every node must be sampled at least once during each cycle, and assuming that~\eqref{eq:beta2} is satisfied, we obtain after some approximations
\begin{equation} \label{eq:tmax}
\theta_{\max} = \max\{{\sigma^2_{\max}}/{(\beta - \sigma^2_{\max})},\;\; 1 \}.
\end{equation}

Following an analogous procedure,  the estimated lower bound $\theta_{\min}$ of $\theta_k$ can be obtained as
\begin{equation} \label{eq:tmin}
\theta_{\min} = \max\{{\sigma^2_{\min}}/{(\beta - \sigma^2_{\min})},\;\; 1 \}.
\end{equation}
Lastly, for $\overline{\theta}_{\max}$ and $\overline{\theta}_{\min}$, we respectively obtain
\begin{equation} \label{eq:t_barramax}
\overline{\theta}_{\max} = \max\{{(\beta - \sigma^2_{\min})}/{\sigma^2_{\min}},\;\; 1 \}
\end{equation}
and
\begin{equation} \label{eq:t_barramin}
\overline{\theta}_{\min} = \max\{{(\beta - \sigma^2_{\max})}/{\sigma^2_{\max}},\;\; 1 \}.
\end{equation}
Thus, using~\eqref{eq:psk}, we can now estimate $p_{s_{\min}}$ and $p_{s_{\max}}$ as 
\begin{equation}	\label{eq:pskquasefinal}
 \widehat{p}_{\min} = \dfrac{\theta_{\min}}{\theta_{\min}+\overline{\theta}_{\max}} 
\end{equation}
and
\begin{equation} \label{eq:pskquasefinal2}
\widehat{p}_{\max} = \dfrac{\theta_{\max}}{\theta_{\max}+\overline{\theta}_{\min}}.
\end{equation}

When $\beta<2\sigma_{\min}^2$, we observe from \eqref{eq:tmin} and \eqref{eq:t_barramax} that 
$\theta_{\min} = \sigma^2_{\min}/{(\beta - \sigma^2_{\min})}$ and  $\overline{\theta}_{\max}=1$. On the other hand, for  $\beta\geq2\sigma_{\min}^2$,~\eqref{eq:tmin} and \eqref{eq:t_barramax} yield 
$\theta_{\min} = 1$ and  $\overline{\theta}_{\max}=
(\beta - \sigma^2_{\min})/\sigma^2_{\min}$, respectively. In both cases, making these replacements in  \eqref{eq:pskquasefinal}, we get
\begin{equation} \label{eq:pskfinal}
\widehat{p}_{\min} = {\sigma_{\min}^2}/{\beta}.
\end{equation}
Analogously, from \eqref{eq:tmax}, \eqref{eq:t_barramin}, and \eqref{eq:pskquasefinal2}
we obtain  
\begin{equation} \label{eq:pskfinal2}
\widehat{p}_{\max} = {\sigma_{\max}^2}/{\beta}.
\end{equation}
Thus, replacing~\eqref{eq:pskfinal} and~\eqref{eq:pskfinal2} in~\eqref{eq:bounds}, we finally get
\begin{equation}
	\label{eq:final}
	V\dfrac{\sigma_{\min}^2}{\beta} \leq \E\{V_s\} \leq V\dfrac{\sigma_{\max}^2}{\beta}.
\end{equation}
For $\beta\!<\!\sigma_{\max}^2$,~\eqref{eq:final} yields an upper bound that is greater than the total number $V$ of nodes, which is not convenient. However, we can generalize it for all $\beta >0$ by recasting it as
\boxedeqn{
	\label{eq:final2}
	V \cdot \min\left\{1,\ \dfrac{\sigma_{\min}^2}{\beta}\right\} \! \leq \! \E\{V_s\} \! \leq \! V \cdot \min\left\{1,\  \dfrac{\sigma_{\max}^2}{\beta}\right\}.
}

Replacing $\beta\!<\!\sigma_{\min}^2$ in~\eqref{eq:final2} implies $\E\{V_s\}\!=\!V$, which agrees with~\eqref{eq:betamin} being a necessary condition to ensure a reduction in the number of sampled nodes. Analogously, replacing $\beta\!>\!\sigma_{\max}^2$ we conclude that $\E\{V_s\}\!<\!V$, which is in accordance with ~\eqref{eq:beta2} being a sufficient condition. 
Moreover, the higher the parameter $\beta$, the smaller the amount of  nodes sampled
in the mean during steady state, as expected.
Since there is a trade-off between the tracking capability and the gains in terms of computational cost provided by the sampling mechanism,
we should care not to  choose excessively high values for $\beta$, since they  can deteriorate the performance in non-stationary environments. Simulation results suggest that if
$\beta\!\leq\!5\sigma_{\max}^2$, the good behavior of the algorithm is maintained.
Moreover, the upper and lower bounds coincide when $\sigma^2_{\min}\! =\! \sigma^2_{\max}$. 
Finally, the step size $\mu_s$ does not affect the number of sampled nodes.

\subsection{Choosing the step size $\mu_s$} \label{sec:mus}

In this section, we show how to
choose a proper value for the parameter $\mu_s$. To do so, we study how fast the nodes cease to be sampled (i.e., how fast we arrive at $\E\{\alpha_k(n)\} \!\leq\! 0$) after the algorithm's initialization with $\alpha_k(0)\!=\!\alpha^+$ for $k\!=\!1,\ \cdots,\ V$. 
From~\eqref{eq:expected_b2} and~\eqref{eq:menorque}, we can write 
\begin{equation} \label{primeiro_passo}
\E\{\Delta \alpha_k(n) \} \leq   \mu_s  \E\{\phi^{\prime}[\alpha_k(n)] \} (\sigma^2_{\max} - \beta).
\end{equation}
Since in this case we consider $\alpha_k(n) \in [0, \alpha^+]$, approximating $\phi^\prime[\alpha_k(n)]$ by its first-order Taylor expansion around $\alpha_k(n) = 0$ is not a suitable approach. Instead, we now approximate $\phi^{\prime}[\alpha_k(n)]$ in that interval by a straight line that crosses the points $(0, \phi^\prime_0)$ and $(\alpha^+\!, \phi^\prime_{\alpha^+})$, in which $\phi^\prime_{0}$ and $\phi^\prime_{\alpha^+}$ respectively denote the value of  $\phi^{\prime}[\alpha_k(n)]$ evaluated at $\alpha_k(n)=0$ and $\alpha_k(n)=\alpha^+$. This approximation is given by
\begin{equation} \label{reta}
\phi^{\prime}[\alpha_k(n)] \approx   \zeta \alpha_k(n)+\phi^\prime_0,
\end{equation}
where $\zeta =  [{\phi^\prime_{\alpha^+}-\phi^\prime_{0}}]/{\alpha^+}$.
For $\alpha^{+}\!=\!4$, this is a good approximation since its mean-squared error in $[0, \alpha^{+}]$ is of the order of $5\times 10^{-4}$. 

Replacing~\eqref{reta} in~\eqref{primeiro_passo}, we obtain
\begin{equation} \label{passo_intermediario}
\E\{ \alpha_k(n+1) \} \lessapprox \E\{ \alpha_k(n) \} (1 + \zeta \rho)  + \phi^\prime_0 \rho,
\end{equation}
where $\rho = \mu_s (\sigma^2_{\max} - \beta)$.
Since we assumed $ \E\{ \alpha_k(n) \}\! \approx\! \alpha^+$ during transient,  we denote the first iteration of the steady state by $n_0$ and define $n_0+\Delta n \triangleq n+1$. Then, considering $ \E\{ \alpha_k(n_0) \}\! \approx\! \alpha^+$ in~\eqref{passo_intermediario} and applying it recursively, we obtain
\begin{equation} \label{segundo_passo}
\E\{\alpha_k(n_0\!+\!\Delta n)\}\! \lessapprox\! \alpha^+ (1 \!+\! \zeta \rho)^{\Delta n}\! +\! \phi^\prime_0 \rho \sum_{\eta=0}^{\Delta n-1} (1 \!+ \!\zeta \rho)^\eta.
\end{equation}
After some algebraic manipulations, we arrive at
\begin{equation} \label{terceiro_passo}
\E\{\alpha_k(n_0\!+\!\Delta n)\} \lessapprox [(\zeta \alpha^+ + \phi^\prime_0) (1 + \zeta \rho)^{\Delta n} - \phi^\prime_0]/{\zeta}.
\end{equation}
Since we are interested in studying how fast we arrive at $\E\{\alpha_k(n)\} \!\!\!\! \leq \!\!\!\! 0$ depending on our choice of $\mu_s$, we set $\E\{\alpha_k(n_0\!+\!\Delta n)\}$ to zero in~\eqref{terceiro_passo}. Thus, for a desired value of $\Delta n$ and $\beta\!>\!\sigma_{\max}^2$, we should choose
\boxedeqn{ \label{eq:ultima_mu}
	\mu_s  > \frac{\alpha^+}{(\beta - \sigma^2_{\max})(\phi_0^\prime - \phi_{\alpha^+}^\prime)} \left[ \left(\frac{\phi_o^\prime}{\phi_{\alpha^+}^\prime} \right)^{\frac{1}{\Delta n}} - 1 \right].
}
From \eqref{eq:ultima_mu}, we observe that the smaller the $\Delta n$, the larger the value of $\mu_s$, which is reasonable.
Moreover, as $\beta$ approaches $\sigma_{\max}^2$,~\eqref{eq:ultima_mu} yields increasingly large values for $\mu_s$. Since~\eqref{eq:beta2} is a sufficient condition, the nodes may cease to be sampled even for $\beta\! \leq\! \sigma_{\max}^2$. When $\beta\! \approx\! \sigma_{\max}^2$ and $\sigma_{\min}^2\!<\!\sigma_{\max}^2$,~\eqref{eq:ultima_mu} may overestimate the value of $\mu_s$ required to cease the sampling of the nodes within $\Delta n$ iterations. Nonetheless, this does not invalidate~\eqref{eq:ultima_mu}, since we are only interested in ensuring that the sampling will cease in at most $\Delta n$ iterations.

\section{Computational cost analysis} \label{sec:cost}

If~\eqref{eq:beta2} is satisfied, the proposed mechanism leads to a reduction in the expected number of sampled nodes. However, this does not necessarily guarantee an advantage in terms of computational cost, since the sampling algorithm also requires a certain number of operations. 
Analyzing Table~\ref{tab:asdnlms}, we see that the sampling mechanism requires $|\mathcal{N}_k|\!+1\!+\!\widebar{s}_k(n)$ sums, $|\mathcal{N}_k|\!+\!1$ multiplications and two comparisons per iteration for each sampled node $k$ of the network. However,  when node $k$ is not sampled, AS-dNLMS does not have to calculate $\xM_k^{\T}(n) \w_k(n)$, $e_k(n)$, and $\mu_k(n)$, thus requiring $2M\!-\!\sum_{i \in \mathcal{N}_k}\!\bar{s}_i(n)$
less multiplications, $2M\!-\!|\mathcal{N}_k|+1$ less sums, and one less division than the original dNLMS. These results are summarized in
Table II for both algorithms with ACW applied to classical distributed signal processing. We should mention that we consider an implementation of $\phi^{\prime}[\alpha_k(n)]$ through a look-up table, which is not taken into account in Table~\ref{tab:cost}.


\begin{table*}[htb!]
	\centering
	\caption{\label{tab:cost} Computational cost comparison between dNLMS and AS-dNLMS with ACW for classical distributed signal processing: number of operations per iteration for each node $k$.}
	\begin{tabular}{|c|c|c|c|c|}
		\hline
		\hline
		Algorithm & Multiplications ($\bigotimes$) & Sums ($\bigoplus$) & Divisions & Comparisons \\
		\hline
		dNLMS & $M(3+|\mathcal{N}_k|)+4$ & $M(3+|\mathcal{N}_k|)+3$ & $|\mathcal{N}_k|$ & 0\\
		\hline
		AS-dNLMS & $\bar{s}_k(n) (2M+2)\!+\!M(1+|\mathcal{N}_k|)\!+\!|\mathcal{N}_k|\!+\!4$ & $\bar{s}_k(n)(2M+2)\!+\!M(|\mathcal{N}_k|+1)\!+\!|\mathcal{N}_k|\!+\!2$ & $|\mathcal{N}_k|\!+\!\widebar{s}_k(n)\!-\!1$ & 2\\
		\hline
		\hline
	\end{tabular}
\end{table*}

In this section, we analyze which conditions have to be satisfied in order to ensure that the computational cost of AS-dNLMS is lower than that of dNLMS. In our analysis, we focus on the number of multiplications ($\otimes$). Analogous results can be obtained for the number of sums, but since they are less restrictive for AS-dNLMS, they are not presented here. 

Firstly, 
we subtract the second row of Table~\ref{tab:cost} from the first one, obtaining
\begin{equation}
\Delta \otimes_k = 2M - 2(M+2)\widebar{s}_k(n) - |\mathcal{N}_k|,
\end{equation}
where $\Delta \otimes_k$ represents the difference in the number of multiplications between dNLMS and AS-dNLMS.

Summing $\Delta \otimes_k$ for $k=1,\cdots,V$ and taking expectations, we obtain for the whole network
\begin{equation} \label{eq:mult1}
\E\{\Delta \otimes\} \!=\! 2VM\! -\sum_{k\!=\!1}^V\!\left[2(M\!+\!2)\E\{\widebar{s}_k(n)\}\!+|\mathcal{N}_k|\right],
\end{equation}
where we have defined $\Delta \otimes \!\triangleq\! \sum_{k=1}^V \!\Delta \otimes_k$.

AS-dNLMS is advantageous over dNLMS in terms of computational cost when $\E\{\Delta \otimes\} > 0 $. Assuming again that $\widebar{s_k}$ can be seen as a Bernoulli random variable in steady state, we have $\E\{\widebar{s}_k(n)\} = p_{s_k}$. In this case, the  worst-case scenario occurs if we consider $\E\{\widebar{s}_k(n)\} = p_{s_{\max}}$, since this minimizes $\E\{\Delta \otimes\}$, leading to
\begin{equation} \label{eq:mult2}
\E\{\Delta \otimes_{\min}\} \!=\! 2V\!M\! -\! \left[2(M-1)V \!+\! \sum_{k=1}^V |\mathcal{N}_k|\right] p_{s_{max}}.
\end{equation}
Enforcing $\E\{\Delta \otimes_{\min}\}\! >\! 0 $, we conclude from~\eqref{eq:mult2} that, in order to ensure that AS-dNLMS requires less multiplications than dNLMS, we must have
\begin{equation} \label{eq:psmax_mult}
p_{s_{\max}} < \dfrac{2VM-\sum_{k=1}^V|\mathcal{N}_k|}{2V(M+2)}.
\end{equation}
Replacing $p_{s_{\max}}$ by $\widehat{p}_{s_{\max}}$ from~\eqref{eq:pskfinal2} in~\eqref{eq:psmax_mult}, we finally~get
\boxedeqn{ \label{eq:beta_mult}
	\beta >  \dfrac{2V(M+2)\sigma_{\max}^2}{2VM-\sum_{k=1}^V|\mathcal{N}_k|}.
}
We remark that~\eqref{eq:beta_mult} is a sufficient (but not necessary) condition to ensure that AS-dNLMS presents a lower computational cost than dNLMS. The factor that multiplies $\sigma_{\max}^2$ in~\eqref{eq:beta_mult} is always greater than one, which is in accordance with our expectations. Moreover,  the right-hand side  of~\eqref{eq:beta_mult} approaches $\sigma^2_{\max}$ as $M$ grows. Thus,  the higher the order of the filter, the greater the computational cost reduction of AS-dNLMS in comparison with dNLMS for  a fixed $\beta$. Furthermore, we can only ensure a decrease in the computational cost if
\boxedeqn{ \label{eq:restricao}
M > \dfrac{\sum_{k=1}^V|\mathcal{N}_k|}{2V}.
}
If~\eqref{eq:restricao} is not satisfied, there is no finite value for $\beta\!>\!0$ that can satisfy the sufficient condition~\eqref{eq:beta_mult}, since this would imply $p_{s_{\max}}\! \leq \!0$ in~\eqref{eq:psmax_mult}. Finally, we remark that we would obtain a different expression for $\beta$ if we considered other diffuse algorithms~\cite{cattivelli2008diffusion,li2009distributed} and other rules for the selection of the combination weights~\mbox{\cite{Sayed_Networks2014,Fernandez-Bes2017,yu2013strategy}.} 


\section{Simulation Results} \label{sec:simulation}

In this section, we present simulation results to illustrate the behavior of the proposed sampling mechanism and to validate the results of Sections~\ref{sec:analysis} and~\ref{sec:cost}. The  results presented were obtained over an average of 100 independent realizations. For the sake of better visualization, we filtered the curves by a moving-average filter with $64$ coefficients.

\begin{figure}[htb!]
		\centering
	\begin{subfigure}[c]{0.48\columnwidth}
			\centering
		\includegraphics*[trim=6cm 10.4cm 5.8cm 10.2cm, clip=true, width=1\columnwidth]{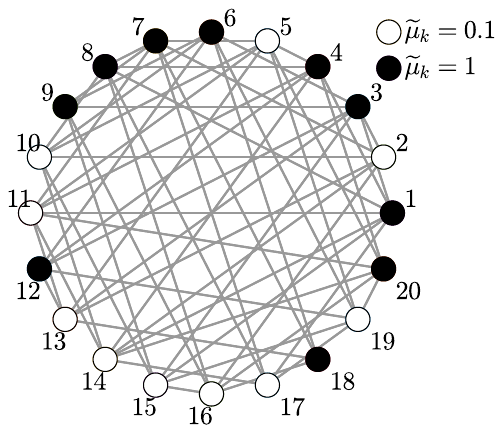}
		\caption{\label{fig:rede}}
	\end{subfigure}
	\begin{subfigure}[c]{0.48\columnwidth}
	\centering
	\includegraphics*[width=1\columnwidth]{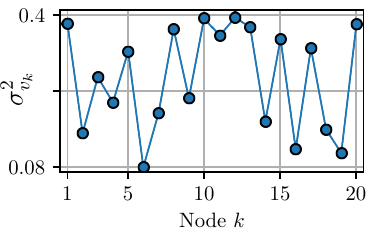}
	\caption{\label{fig:ruido}}
\end{subfigure}
\caption{\label{fig:rede_ruido}(a) Network used in the simulations. Nodes represented by filled circles employ ATC dNLMS with $\widetilde{\mu}_k \!=\! 1$, whereas nodes represented by empty circles run the same algorithm with $\widetilde{\mu}_k \!=\! 0.1$. (b) Noise variance $\sigma_{v_k}^2$ for $k=1,\cdots,V$.}
\end{figure}

We consider the ATC dNLMS algorithm and a heterogeneous network with 20 nodes. Half of the them use $\widetilde{\mu}_k = 0.1$, while the other half uses $\widetilde{\mu}_k = 1$, as depicted in Fig.~\ref{fig:rede_ruido}(a). Furthermore, each node $k$ is subject to a different noise variance $\sigma_{v_k}^2$, as shown in Fig.~\ref{fig:rede_ruido}(b). For the optimal system $\w^{\rm o}$, we consider a random vector with $M\!=\!50$ coefficients uniformly distributed in $[-1,1]$.

The combination weights are updated using the ACW algorithm with $\nu_k\!=\!0.2$ for $k\!=\!1,\cdots,V$~\cite{tu2011optimal}, and we use $\delta \!=\!\delta_c\!=\! 10^{-5}$ as regularization factors. As a performance indicator, we adopt the network mean-square-deviation (NMSD), given by
\begin{equation}
\text{NMSD}(n) = 	\frac{1}{V} \textstyle\sum_{k=1}^V \E \{ \norm{\w^{\rm o}(n) \!-\! \w_k(n)}^2\}.
\end{equation}
Moreover, in some situations we also analyze the network mean-square-error (NMSE), given by
\begin{equation}
\text{NMSE}(n) = 	\frac{1}{V} \textstyle\sum_{k=1}^V \E \{ e_k^2(n)\}.
\end{equation}	

For the ease of understanding, this section is divided as follows. In Subsection~\ref{sec:comparacao_com_aleatorio}, we compare AS-dNLMS with the random sampling technique of Fig.~\ref{fig:intro}. The theoretical results of Section~\ref{sec:analysis} are validated in Subsection~\ref{sec:validation}, and in~\ref{sec:application_censoring} we compare ASC-dNLMS to other censoring techniques. Next, in Subsection~\ref{sec:track}, we study the tracking capability of the proposed techniques. Finally, in~\ref{sec:application_censoring}, we employ AS-dNLMS in the context of graph distributed adaptive filtering.
	
\subsection{Comparison with Random Sampling} \label{sec:comparacao_com_aleatorio}

\begin{figure}[htb!]
	\centering
	\includegraphics*[trim=0cm 0.2cm 0cm 0.1cm, clip=true,width=0.8\columnwidth]{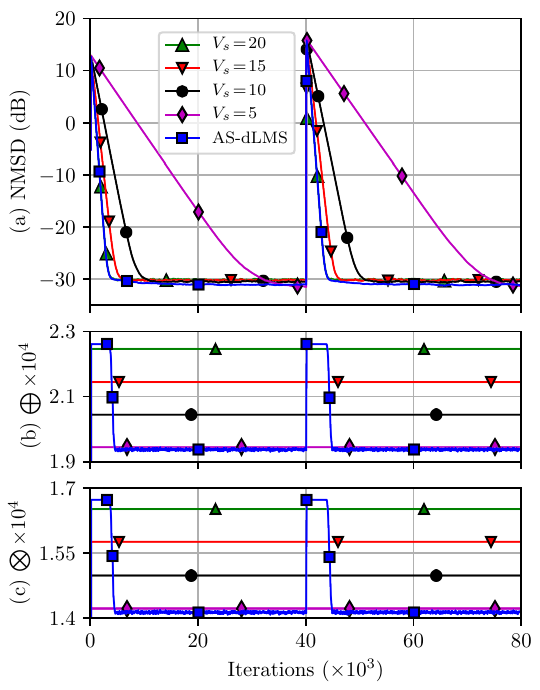}
	\caption{\label{fig:compara_com_aleatorio} Comparison between dNLMS with  a random sampling technique with different amounts of sampled nodes and AS-dNLMS (\mbox{$\beta \!=\! 1.6 \sigma_{\max}^2 \!=\! 0.64$}, \mbox{$\mu_s \! = \! 0.06$}). (a) NMSD curves, (b) Sums, and (c) Multiplications per iteration.}
\end{figure}

Firstly, we return to the simulation of Fig.~\ref{fig:intro} and compare the behavior of AS-dNLMS to that of the original dNLMS with the random sampling technique
and different numbers of sampled nodes $V_s$. Nonetheless, here we simulate a change in the environment by flipping the parameter vector $\w^{\rm o}$ in the middle of each realization. For the network of Fig.~\ref{fig:rede_ruido},~\eqref{eq:restricao} yields $M\!>\!5.4$, which is thus satisfied. For $M\!=\!50$,~\eqref{eq:beta_mult} in its turn yields $\beta \!>\! 1.0610 \sigma_{\max}^2$. We adjusted  AS-dNLMS  to obtain approximately the same computational cost as that of dNLMS with $V_s\!=\!5$ nodes sampled. For this purpose, we adopted $\beta=1.6\sigma_{\max}^2$ and $\mu_s=0.06$. 
Figs.~\ref{fig:compara_com_aleatorio}(a),~\ref{fig:compara_com_aleatorio}(b) and~\ref{fig:compara_com_aleatorio}(c)
present respectively the NMSD performance and the average number of sums and multiplications per iteration.
As seen in Fig.~\ref{fig:intro}, the more nodes are sampled during the transient, the faster the convergence rate. Moreover, we observe that AS-dNLMS is able to detect the change in the optimal system and, since all nodes are sampled during the transients, it converges
as fast as the dNLMS algorithm with all nodes sampled.
From Figs.~\ref{fig:compara_com_aleatorio}(b) and~\ref{fig:compara_com_aleatorio}(c) we observe that during the transients the computational cost of AS-dNLMS is slightly higher than that of the dNLMS algorithm with all nodes sampled, as expected, but decreases significantly in steady-state.

\subsection{Validation of the Theoretical Analysis} \label{sec:validation}

\begin{figure}[htb!]
	\centering
	\includegraphics*[width=0.9\columnwidth]{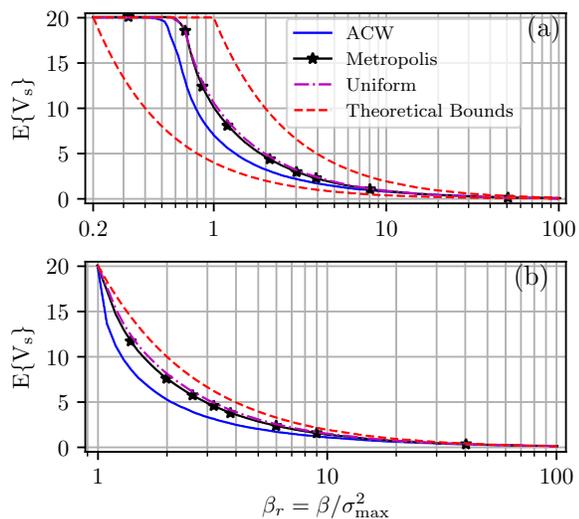}
	\caption{\label{fig:prev_beta} Theoretical bounds and average number of nodes sampled by AS-dNLMS with three combination rules as a function of $\beta\!\!\geq\!\!\sigma_{\min}^2$. (a) $\sigma_{v_k}^2$ as in Fig.~\ref{fig:rede_ruido}. (b) $\sigma_{v_k}^2\!=\!0.4$ for $k\!=\!1,\cdots,V$.}
\end{figure}

In order to validate~\eqref{eq:final2}, we also tested the AS-dNLMS algorithm in a stationary environment with different values of $\beta \geq \sigma_{\min}^2$ and three methods for the selection of the combination weighs: the Uniform and Metropolis rules~\cite{Sayed_Networks2014}, and the ACW algorithm~\cite{tu2011optimal}. Two scenarios were considered: one with the noise power in the network distributed as in Fig.~\ref{fig:ruido}, and another where $\sigma_{v_k}^2 = 0.4$ for $k=1,\cdots,V$. The results are shown in Fig.~\ref{fig:prev_beta}(a) and~\ref{fig:prev_beta}(b), respectively. For the ease of visualization, they are presented in terms of $\beta_r$, defined as $\beta_r \triangleq {\beta}/{\sigma_{\max}^2}.$ Along with the experimental data, the predicted upper and lower bounds $\widehat{V}_{s_{\max}}$ and $\widehat{V}_{s_{\min}}$ are presented for each $\beta_r$ using dashed lines.  We should notice that these bounds coincide in Fig.~\ref{fig:prev_beta}(b), since $\sigma_{\min}^2\!=\!\sigma_{\max}^2$ in this case. Moreover, in Fig.~\ref{fig:prev_beta}(a), the upper bound remains fixed at $V_{s_{\max}}\!=\!V\!=\!20$ for $\beta\leq\sigma_{\max}^2$. We also observe from Fig.~\ref{fig:prev_beta} that the higher $\beta$ is, the less nodes are sampled in both scenarios, as expected.
Furthermore, the experimental data 
lie between the theoretical bounds for all combination rules and for all values of $\beta_r$ in Fig~\ref{fig:prev_beta}(a). On the other hand, from~\ref{fig:prev_beta}(b) we notice that the theoretical model slightly overestimates the number of sampled nodes for $1\!<\!\beta_r\!\leq\!20$. In both cases, the adoption of the ACW algorithm led to a smaller number of sampled nodes in comparison with the Uniform and Metropolis rules. 

\begin{figure}[htb!]
	\centering
	\includegraphics*[trim=0cm 0.2cm 0cm 0.1cm, clip=true,width=0.8\columnwidth]{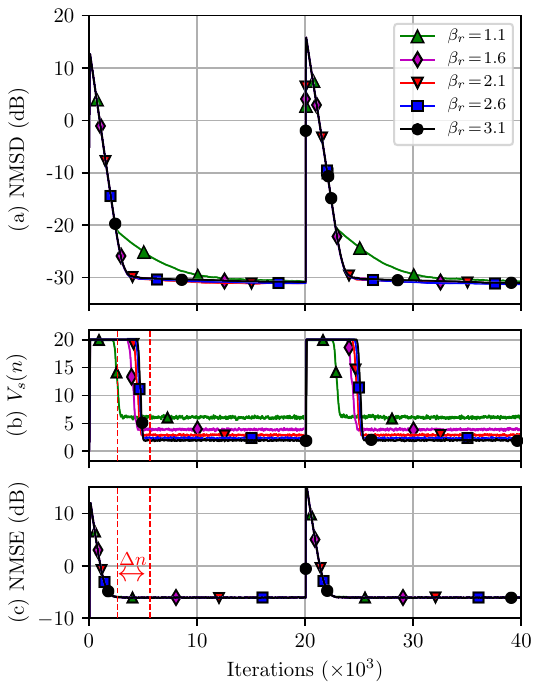}
	\caption{\label{fig:mu1} Simulation results obtained with $1.1 \sigma_{\max}^2 \!\leq\! \beta\! \leq\! 3.1 \sigma_{\max}^2$ and $\mu_s$ adjusted by~\eqref{eq:ultima_mu} for each case. (a) NMSD curves, (b) Number of sampled nodes per iteration, and (c) NMSE Curves.}
\end{figure}

In Fig.~\ref{fig:mu1}, we test~\eqref{eq:ultima_mu} by using it to set the step size $\mu_s$ for different values of $\beta$ with $\Delta n \! = \! 3000$. In Fig.~\ref{fig:mu1}(a) we show the NMSD curves, in Fig.~\ref{fig:mu1}(b) the number of sampled nodes per iteration, and in Fig.~\ref{fig:mu1}(c) the NMSE.

From Figs.~\ref{fig:mu1}(b) and~\ref{fig:mu1}(c) we observe that, before the abrupt change in the optimal system, the number of sampled nodes stabilizes at approximately the same time for all $\beta_r\!>\!1.1$. For $\beta_r\!=\!1.1$, we can notice that~\eqref{eq:ultima_mu} slightly overestimates $\mu_s$. This is expected for $\beta_r\gtrapprox1$, as discussed in Section~\ref{sec:mus}. In this case, AS-dNLMS ceased to sample the nodes before reaching the steady state in terms of NMSD, which compromised the convergence rate. This illustrates the importance of a proper choice for $\mu_s$ as well as $\beta$. Nonetheless, since the sampling of the nodes ceased in less than $\Delta n$ iterations after the beginning of the steady state in terms of NMSE, the results obtained support the validity of~\eqref{eq:ultima_mu}. However, this shows that some care must be taken when using~\eqref{eq:ultima_mu} for~$\beta\gtrapprox\sigma_{\max}^2$. 


\begin{figure}[htb!]
	\centering
	\includegraphics*[trim=0cm 0.2cm 0cm 0.1cm, clip=true,width=0.8\columnwidth]{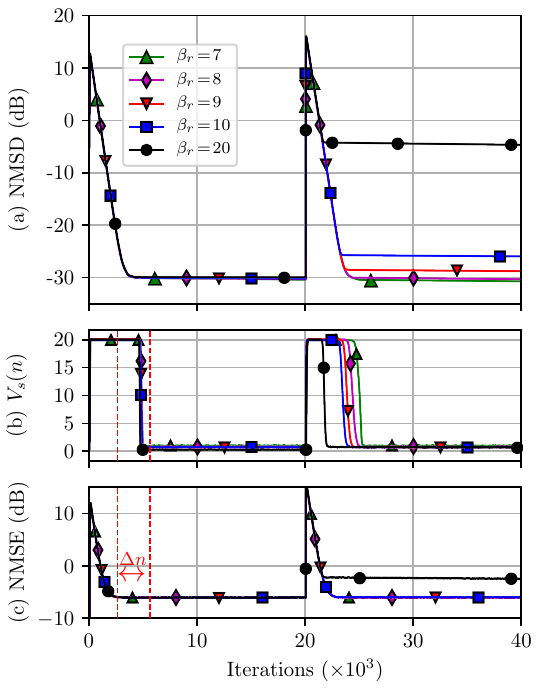}
	\caption{\label{fig:mu2} Simulation results obtained with $7 \sigma_{\max}^2 \! \leq \! \beta \! \leq \! 20 \sigma_{\max}^2$ and $\mu_s$ adjusted by~\eqref{eq:ultima_mu} for each case. (a) NMSD curves, (b) Number of sampled nodes per iteration, and (c) NMSE Curves.}
\end{figure}

In Fig.~\ref{fig:mu2} we repeated the experiments of Fig.~\ref{fig:mu1} with higher values of $\beta_r$. We observe that the number of sampled nodes stabilizes almost simultaneously for all values of $\beta_r$ before the abrupt change and that the performance of AS-dNLMS is maintained before the change in the optimal system. Nonetheless, after the change occurs, the NMSD is affected for $\beta_r\geq 8$. The higher the parameter $\beta$, the more intense the deterioration in performance. The difference in the behavior of the algorithm before and after the change in the optimal system can be explained by the initialization with $\alpha_k(0) = \alpha^+$ for $k=1,\cdots,V$. In contrast, right before the abrupt change, we have $\alpha_k(n) \ll \alpha^+$. Thus, the algorithm ceases to sample the nodes earlier in this case, as can be seen in Fig.~\ref{fig:mu2}(b). We recall that $\beta\leq5 \sigma_{\max}^2$ seems to be a safe interval for the choice of $\beta$, according to various simulations results. 


\subsection{Application as a Censoring Technique} \label{sec:application_censoring}

\begin{figure}[htb!]
	\centering
	\includegraphics*[trim=0cm 2.cm 0cm 0.2cm, clip=true,width=0.8\columnwidth]{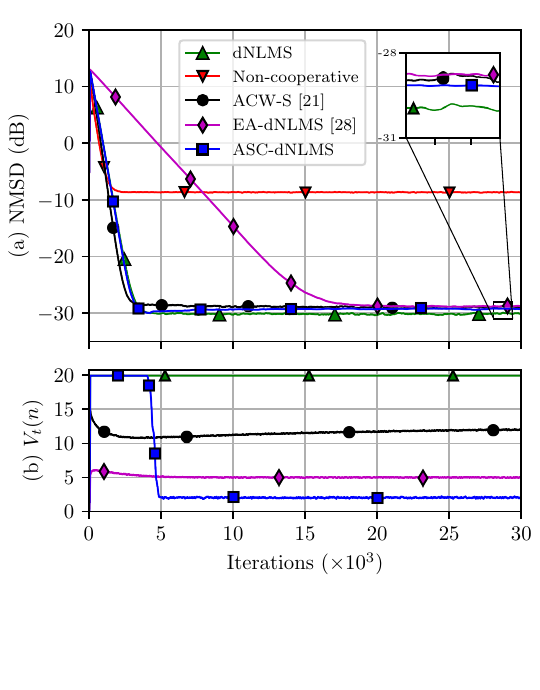}
	\caption{\label{fig:censoring} Comparison between the ASC-dNLMS, ACW-S and EA-dNLMS  algorithms. The parameters adopted are shown in Table~\ref{tab:param}. (a) Steady-state NMSD curves. (b) Number of broadcasts per iteration. }
\end{figure}

In this section, we test the ASC-dNLMS algorithm and compare it to other techniques found in the literature, namely, the ACW-Selective (ACW-S) algorithm of~\cite{arroyo2013censoring} and the energy-aware diffusion algorithm (EA-dNLMS) of~\cite{gharehshiran2013distributed}. Assuming that the nodes can broadcast their data to all of their neighbors at once, we present in Fig.~\ref{fig:censoring}(a) the NMSD curves, and in Fig.~\ref{fig:censoring}(b), the number $V_t(n)$ of transmitting nodes per iteration, i.e. the amount of broadcasts in the network. 

The algorithms were adjusted to achieve approximately the same level of steady-state NMSD. Table~\ref{tab:param} shows the adopted values for the parameters of each solution. In this regard, it is worth noting that EA-dNLMS presents a high number of parameters, which may be difficult to adjust. We consider the version of EA-dNLMNS that allows node $k$ to receive and combine the estimates from its neighbors even when it is not transmitting~\cite{gharehshiran2013distributed}, and we adopt a normalized step size following~\eqref{eq:muk}. For comparison, we also present results obtained with the original dNLMS and with the non-cooperative case.

Unlike AS-dNLMS, which maintained the steady-state performance of dNLMS, ASC-dNLMS achieves a slightly higher level of NMSD in steady state in comparison with the original algorithm. The same occurs for the ACW-S and EA-dNLMS algorithms, as can be seen in Fig.~\ref{fig:censoring}(a). We observe that EA-dNLMS presents a notably slower convergence rate in comparison with ACW-S and ASC-dNLMS, which converge at a rate similar to that of dNLMS. On the other hand, from Fig.~\ref{fig:censoring} we see that ACW-S utilizes a comparatively high number of broadcasts, thus saving less energy. During steady state, both ACW-S and EA-dNLMS transmit more  than the proposed ASC-dNLMS, which maintains all transmissions during the transient but drastically reduces the number of broadcasts after converging. Thus, the proposed technique saves more energy in steady state while preserving the convergence rate.

\begin{table}[h!tb]
	\centering
	\caption{Parameters of the algorithms used in the simulations of Fig.~\ref{fig:censoring}} \label{tab:param}
	\begin{tabularx}{\columnwidth}{|c|X|}
		\hline
		ACW-S~\cite{arroyo2013censoring} & $E_T\!=\!1$, $E_T\!=\!2$ \\
		\hline
		EA-dNLMS~\cite{gharehshiran2013distributed} & $E_{\text{Act}}\!=\!33.5966 \cdot 10^{-3}$, $E_{\text{Tx}}\!=\!15.16 \cdot 10^{-3}$, $K_{\ell,1}\!=\!2$, $K_{\ell,2}\!=\!0.5$, $K_{g}\!=\!2$, $\gamma_g\!=\!2$,$\gamma_\ell\!=\!2$, $\delta \!=\! 0.5$, $\rho\!=\!0.01$, $r\!=\!2$\\
		\hline
		ASC-dNLMS & $\beta\! =\! 2.1\sigma_{\max}^2$, $\mu_s\! =\! 0.0333$ \\
		\hline
	\end{tabularx}
	
\end{table}

\subsection{Random-Walk Tracking} \label{sec:track}
\begin{figure}[htb!]
	\centering
	\includegraphics*[trim=0cm 0.2cm 0cm 0.1cm, clip=true,width=0.8\columnwidth]{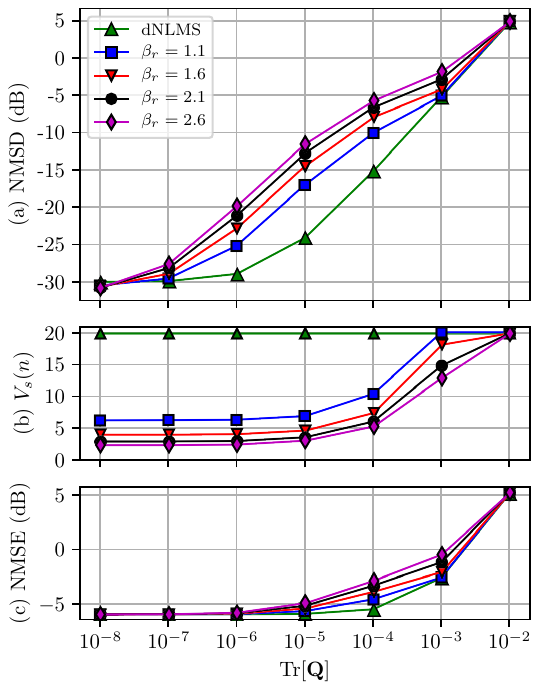}
	\caption{\label{fig:random_walk} Simulation results in a nonstationary environment following Model~\eqref{eq:nonstationary}. (a) Steady-state NMSD, (b) Number of nodes sampled per iteration, and (c) Steady-state NMSE. }
\end{figure}

As can be observed from Fig.~\ref{fig:mu2}, increased values of $\beta$ may hinder the tracking capability of AS-dNLMS. Thus, in this section, we investigate the behavior of the algorithm in nonstationary environments following a random-walk model, in which the optimal solution $\w^{\rm o}(n)$ varies according to
\begin{equation} \label{eq:nonstationary}
	\w^{\rm o}(n) = \w^{\rm o}(n-1)+\mathbf{q}(n),
\end{equation}
where $\mathbf{q}(n)$ is a zero-mean i.i.d.  column vector with length $M$ and autocovariance matrix $\mathbf{Q} = E\{\mathbf{q}(n)\mathbf{q}^{\rm T}(n)\}$ independent of any other signal. This model is commonly used in the adaptive filtering literature~\cite{SayedL2008,Fernandez-Bes2017}. In our experiments, we consider a Gaussian distribution for $\mathbf{q}(n)$ with $\mathbf{Q} = \sigma_{q}^2 \mathbf{I}$, where $\mathbf{I}$ denotes the identity matrix. 
In Fig.~\ref{fig:random_walk}, we present the results obtained with the AS-dNLMS algorithm and different values of $\beta$ as a function of $\Tr[\mathbf{Q}]$. For each $\beta_r$, we maintained the corresponding step size $\mu_s$ used in the simulations of Fig.~\ref{fig:mu1}. For comparison, we also show the results obtained with the dNLMS algorithm with all nodes sampled. In Fig.~\ref{fig:random_walk}(a), we present the steady-state levels of NMSD, in Fig.~\ref{fig:random_walk}(b) the average number of sampled nodes per iteration and in Fig.~\ref{fig:random_walk}(c) the steady-state NMSE. The results presented were obtained by averaging the data over the last 600 iterations of each realization, after all the algorithms achieved steady state.

From Fig.~\ref{fig:random_walk}(a) we can observe that, in slowly-varying environments ($\Tr[\mathbf{Q}] = 10^{-8}$), the performance of AS-dNLMS is similar to that of dNLMS with all nodes sampled. However, for $10^{-7} \leq \Tr[\mathbf{Q}] \leq 10^{-3}$, there is a degradation in performance in comparison with dNLMS. The higher the parameter $\beta$, the more intense this deterioration becomes for a fixed value of $\Tr[\mathbf{Q}]$. For $\Tr[\mathbf{Q}] \leq 10^{-5}$ and a fixed $\beta$, this deterioration in comparison with dNLMS intensifies with the increase of $\Tr[\mathbf{Q}]$. On the other hand,  for $\Tr[\mathbf{Q}] > 10^{-5}$, the difference in performance begins to decrease as the variations in the optimal system become faster. This can be explained by analyzing Figs.~\ref{fig:random_walk}(b) and~\ref{fig:random_walk}(c). We observe that, when the environment varies slowly or moderately, the number of nodes sampled by the AS-dNLMS is not significantly affected by the increase of $\Tr[\mathbf{Q}]$. This occurs since the effects of the changes in the optimal system are small in comparison with those of the measurement noise for $\Tr[\mathbf{Q}]<10^{-5}$, and thus the NMSE does not increase noticeably, as seen in Fig.~\ref{fig:random_walk}(c). However, as these variations become faster, they begin to affect the estimation error more intensely, and the NMSE starts to increase for $\Tr[\mathbf{Q}]\geq 10^{-5}$, leading to a gradual rise in the number of sampled nodes in Fig.~\ref{fig:random_walk}(b). For $\Tr[\mathbf{Q}] = 10^{-2}$, the algorithm does not cease to sample any of the nodes for $\beta_r \leq 2.6$, and thus its performance matches that of dNLMS. 



\begin{figure}[t]
	\centering
	\includegraphics*[trim=0cm 2.cm 0cm 0.1cm, clip=true,width=0.8\columnwidth]{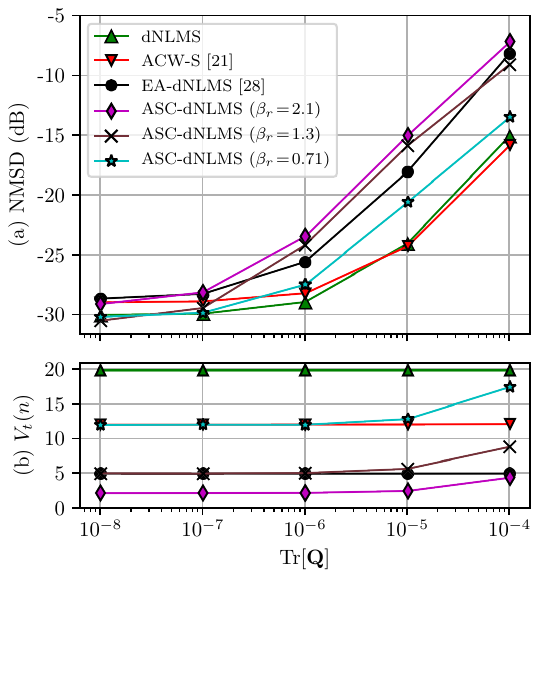}
	\caption{\label{fig:random_walk2} Simulation results in a nonstationary environment following Model~\eqref{eq:nonstationary} with the algorithms listed in Table~\ref{tab:param}. (a) Steady-state NMSD, and (b) Broadcasts per iteration. }
\end{figure}

Next, we repeated the experiment of Fig.~\ref{fig:random_walk} for \mbox{ASC-dNLMS}, ACW-S and EA-dNLMS with the parameters of Table~\ref{tab:param}. The results are shown in Fig.~\ref{fig:random_walk2}. We also present the results obtained with ASC-dNLMS with $\beta_r = 1.3$ and $\beta_r = 0.71$, which were respectively adjusted to lead to the same number of broadcasts as those of EA-dNLMS and \mbox{ACW-S} for $\Tr[\mathbf{Q}]\leq 10^{-6}$. Finally, we also show results obtained with the dNLMS algorithm. We observe from Fig.~\ref{fig:random_walk2}(a) that ASC-dNLMS with $\beta_r=2.1$ achieves a performance similar to that of the other solutions for $\Tr[\mathbf{Q}]=10^{-8}$ and $\Tr[\mathbf{Q}]=10^{-7}$. However, it is outperformed for $\Tr[\mathbf{Q}]\geq 10^{-6}$. It also employs less transmissions than any other solution in these scenarios. With $\beta_r=1.3$, ASC-dNLMS outperforms EA-dNLMS for  $\Tr[\mathbf{Q}]\leq 10^{-7}$ and  $\Tr[\mathbf{Q}] = 10^{-4}$, although its NMSD is higher for $\Tr[\mathbf{Q}] = 10^{-6}$ and  $\Tr[\mathbf{Q}] = 10^{-5}$. With $\beta_r=0.71$, ASC-dNLMS outperforms ACW-S for $\Tr[\mathbf{Q}] \leq 10^{-7}$, while the opposite occurs for other values of $\Tr[\mathbf{Q}]$. The results suggest that ASC-dNLMS generally outperforms ACW-S and EA-dNLMS in stationary or slowly-varying environments while utilizing the same number of transmissions. Moreover, in these cases it can achieve a comparatively similar performance while transmitting less. However, ASC-dNLMS must be employed with caution in scenarios in which the optimal system changes rapidly. Finally, we observe that it is possible to control the trade-off between energy saving and performance by adjusting $\beta$.

\subsection{Application in Graph Adaptive Filtering} \label{sec:graph_filter}

Finally, in this section we employ the proposed sampling algorithm in a graph diffuse adaptive filter. We still consider the network of Fig.~\ref{fig:rede_ruido}(a), and we use its unweighted adjacency matrix, normalized by its largest eigenvalue, as the graph shift operator. Moreover, we consider a scaled version of the noise power distribution of Fig.~\ref{fig:rede_ruido}(b) so as to maintain the same average signal-to-noise ratio (SNR) as before. This resulted in $0.0018 \!\leq\! \sigma_{v_k}^2 \!\leq\! 0.009$ for $k\!=\!1,\cdots,V$. For the optimal system $\w^{\rm o}$, we consider a random vector with $M\!=\!10$ coefficients uniformly distributed in $[-1,1]$, and we flip it in the middle of each realization to simulate a change in the environment. We remark that~\eqref{eq:beta_mult} and~\eqref{eq:restricao} do not hold for graph dNLMS and AS-dNLMS, since they require more operations than their counterparts for classical distributed signal processing.

\begin{figure}[htb!]
	\centering
	\includegraphics*[trim=0cm 0.2cm 0cm 0.1cm, clip=true,width=0.8\columnwidth]{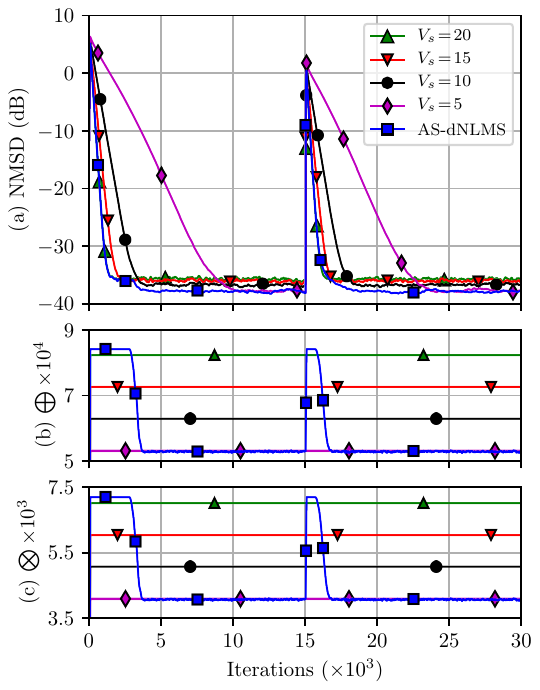}
	\caption{\label{fig:simu_graph} Comparison between graph  dNLMS with  a random sampling technique with different amounts of sampled nodes and AS-graph dNLMS (\mbox{$\beta \!=\! 1.8 \sigma_{\max}^2 \!=\! 0.64$}, \mbox{$\mu_s=2.0364$}). (a) NMSD curves, (b) Sums, and (c) Multiplications per iteration.}
\end{figure}

We adjusted AS-dNLMS to present approximately the same computational cost as that of dNLMS with $V_s\!=\!5$ nodes sampled. Thus, we adopted $\beta=1.8\sigma_{\max}^2$ and $\mu_s=2.0364$, which was obtained by using~\eqref{eq:ultima_mu} with $\Delta n \! = \! 3000$. 
Figs.~\ref{fig:simu_graph}(a),~\ref{fig:simu_graph}(b) and~\ref{fig:simu_graph}(c)
present respectively the NMSD performance and the average number of sums and multiplications per iteration.
Again, we see that  AS-dNLMS detects the change in the optimal system and, converges
as fast as the dNLMS algorithm with all nodes sampled, since it maintains the sampling of all the nodes during the transients. From Figs.~\ref{fig:compara_com_aleatorio}(b) and~\ref{fig:compara_com_aleatorio}(c) we see that during the transients its computational cost is slightly higher than that of the dNLMS algorithm with all nodes sampled, but decreases drastically during steady-state. Comparing Figs.~\ref{fig:compara_com_aleatorio} and~\ref{fig:simu_graph}, we observe that the adaptive sampling mechanism behaves similarly when applied to graph adaptive filtering or to classical distributed signal processing.

\section{Conclusions and Future Work} \label{sec:conclusoes}

In this paper, we proposed adaptive mechanisms for sampling and censoring over distributed solutions. The resulting algorithms, respectively named as AS-dNLMS and ASC-dNLMS, use the information from more nodes when the error in the network is high and from less nodes otherwise. They feature fast convergence rates while significantly reducing the computational cost and the consumption of energy associated with the communication between nodes. Furthermore, we derived analytical expressions that help understand the roles of the parameters $\beta$ and $\mu_s$ and their effects in terms of performance, computational cost reduction, and energy saving. These theoretical results allow to choose proper values for $\beta$ and $\mu_s$ and were validated by the simulation results. It was shown that AS-dNLMS maintains the performance of the original diffuse NLMS algorithm, while noticeably reducing the computational burden. Moreover, it can be employed in graph adaptive filtering as well as in classical distributed signal processing. It was also shown that ASC-dNLMS is capable of saving more energy than other state-of-the-art techniques while achieving a similar steady-state performance and preserving the convergence rate of dNLMS. We should notice that the proposed techniques must be employed with some caution in rapidly-varying environments, as their performance may deteriorate in comparison with dNLMS or other techniques. In their current form, this seems to be the main limitation of the proposed algorithms. For future work, we intend to improve their tracking capability of in such scenarios. Nevertheless, encouraging results for slowly-varying environments indicate AS-dNLMS and ASC-dNLMS as the recommended solutions for such cases, in which they outperform similar techniques. Finally, the proposed mechanisms could also be used on other distributed solutions, such as diffuse recursive least-squares~\cite{cattivelli2008diffusion} or the diffuse affine projection algorithm~\cite{li2009distributed}, which is another suggestion for future research.

\appendices
\section{Deriving Equations~\eqref{eq:tmax} to~\eqref{eq:t_barramin}} \label{sec:demonstracao}

In order to estimate upper or lower bounds for $\theta_k$, we must understand under which circumstances node $k$ remains sampled for the greatest (or lowest) number of iterations in the mean. This can be achieved by estimating the maximum and minimum values $\E\{\alpha_k(n)\}$ and $\E\{\Delta \alpha_k(n)\}$ can assume in the mean during steady state when node $k$ is sampled (i.e., $\widebar{s}_k\!=\!1$). Performing the same analysis for $\widebar{s}_k\!=\!0$, we can determine upper and lower bounds for $\overline{\theta}_k$. For simplicity, we assume in our calculations that~\eqref{eq:beta2} is satisfied, although the final result is generalized in Section~\ref{sec:number} for all $\beta\!>\!0$.

Firstly, let us assume that at a certain iteration $n$, $\alpha_k(n)$ is negative but close to zero.
Setting $\alpha_k(n)$ to zero in~\eqref{eq:b_modificada} and taking expectations, we obtain
\begin{equation} \label{eq:limite1}
\E\{\alpha_k(n+1)\} = \mu_s \phi_0^\prime \textstyle\sum_{i \in \mathcal{N}_k}\!\!c_{ik}(n) \E \{\varepsilon_i^2(n) \},
\end{equation}
Thus, at $n\!+\!1$ the sampling of node $k$ resumes and,
recalling~\eqref{eq:beta2}, $\E\{\Delta \alpha_k(n\!+\!1)\}\!<\!0$. Therefore, from iteration $n\!+\!1$ onwards,
$\alpha_k$ decreases until it becomes negative again, meaning that~\eqref{eq:limite1} yields the maximum value $\alpha_k$ can assume in the mean in steady state. Moreover, assuming~\eqref{eq:menorque2},~\eqref{eq:limite1} yields a different value for each node $k$ that lies~in
\begin{equation}
\mu_s \phi_0^\prime \sigma^2_{\min} \leq \E\{\alpha^{\text{s.s.}}_{k_{\max}}\} \leq \mu_s \phi_0^\prime \sigma^2_{\max},
\end{equation}
where $\E\{\alpha^{\text{s.s.}}_{k_{\max}}\}$ denotes the maximum value $\alpha_k(n)$ can assume in the mean in steady state. Analogously, we now assume that at a certain iteration $n$, $\alpha_k(n)$ is positive but approximately zero. Making this replacement in~\eqref{eq:b_modificada} and taking expectations, we obtain
\begin{equation} \label{eq:limite2}
\E\{\alpha_k(n+1)\} = \mu_s \phi_0^\prime \E \left\{\textstyle\sum_{i \in \mathcal{N}_k}\!\!c_{ik}(n) \varepsilon_i^2(n) -\beta \right\}.
\end{equation}
Since $\alpha_k(n)<0$ and $\E\{\Delta \alpha_k(n)\}>0$ while node $k$ is not being sampled,~\eqref{eq:limite2} provides the minimum value $\alpha_k$ can assume in the mean during steady state. For each node $k$,~\eqref{eq:limite2} yields a different value that lies in the interval
\begin{equation}
\mu_s \phi_0^\prime (\sigma^2_{\min}-\beta) \leq \E\{\alpha^{\text{s.s.}}_{k_{\min}}\} \leq \mu_s \phi_0^\prime (\sigma^2_{\max}-\beta),
\end{equation}
where $\E\{\alpha^{\text{s.s.}}_{k_{\min}}\}$ denotes the minimum value $\alpha_k(n)$ can assume in the mean in steady state.

Since $\E\{\alpha_k(n)\}$ keeps oscillating around the point $\E\{\alpha_k(n)\}\!=\!0$ during steady state, we replace $\phi^\prime[\alpha_k(n)]$ in~\eqref{eq:expected_b2} by its first-order Taylor expansion around $\alpha_k(n)=0$, which is simply equal to the constant $\phi^\prime_0$. 
Thus, when node $k$ is being sampled ($\widebar{s}_k =1$),
subtracting $\alpha_k(n)$ from both sides of~\eqref{eq:expected_b2} and taking expectations yields
\begin{equation} \label{eq:delta_alpha1}
\!-\mu_s \phi_0^\prime   (\beta \!-\! \sigma^2_{\min} ) \! \leq \!\!  \E\{\Delta \alpha_k(n)\} \!\! \leq \!  -\mu_s \phi_0^\prime   (\beta \!-\! \sigma^2_{\max} )\!<\!0.
\end{equation}
Analogously, when the node is not sampled ($\widebar{s}_k =0$),
\begin{equation} \label{eq:delta_alpha2}
\mu_s \phi_0^\prime  \sigma^2_{\min} \! \leq \!  \E\{\Delta \alpha_k(n)\} \! \leq \!  \mu_s \phi_0^\prime \sigma^2_{\max}.
\end{equation}
Thus, in both cases there are upper and lower bounds for $\E\{\Delta \alpha_k(n)\}$ during steady state.

From a certain iteration $n_0$ onward, we consider the model
\begin{equation}
\E\{\alpha_k(n_0+\theta_{k})\}\!=\!\E\{\alpha_k(n_0)\}+\theta_{k}\E\{\Delta\alpha_k(n)\}.\label{eq:geral}
\end{equation}
In order to estimate an upper bound $\theta_{\max}$ for $\theta_k$, we assume that
$\E\{\alpha_k(n_0)\}\!=\!\E\{\alpha^{\text{s.s.}}_{k_{\max}}\}$ and calculate the expected number of iterations required for $\E\{\alpha_k(n)\}$ to fall
below zero in the
scenario where the node is sampled for the maximum number of iterations.
This occurs if  $\E\{\alpha_k(n_0)\}\! =\! \mu_s \phi_0^\prime \sigma^2_{\max}$, which is the upper bound for
$\E\{\alpha^{\text{s.s.}}_{k_{\max}}\}$, and $\E\{\Delta \alpha_k(n)\}\! =\! -\mu_s \phi_0^\prime (\beta - \sigma^2_{\max})$,
which is the least negative variation for $\E\{\Delta\alpha_k(n)\}$ according to~\eqref{eq:delta_alpha1}.
Making $\theta_k\!=\!\theta_{\max}$, setting $\E\{\alpha_k(n_0+\theta_{\max})\}\!=\!0$ in \eqref{eq:geral}, and taking into account the fact that the node must be sampled at least once during each cycle, after some algebra we obtain~\eqref{eq:tmax}. Analogously,
using \eqref{eq:geral} for the lower bound $\theta_k\!=\!\theta_{\min}$, we get~\eqref{eq:tmin}.

For $\overline{\theta}_k$, we replace $\theta_k$ in \eqref{eq:geral} by $\overline{\theta}_k$ and consider that at the iteration $n_0$,
$\E\{\alpha_k(n_0)\}=\E\{\alpha^{\text{s.s.}}_{k_{\min}}\}$.
Thus, the upper bound $\overline{\theta}_{\max}$ for $\overline{\theta}_k$
can be obtained by setting
$\E\{\alpha_k(n_0)\}\! =\! \mu_s \phi_0^\prime \sigma^2_{\min}$,
which is the lower bound for $\E\{\alpha^{\text{s.s.}}_{k_{\min}}\}$, and $\E\{\Delta \alpha_k(n)\}\! =\! \mu_s \phi_0^\prime \sigma^2_{\min}$,
which is the minimum value for $\E\{\Delta\alpha_k(n)\}$ according to~\eqref{eq:delta_alpha2}. Thus,~\eqref{eq:t_barramax} is obtained.
Finally, as an estimate for the lower bound $\overline{\theta}_{\min}$ of $\overline{\theta}_k$, we get~\eqref{eq:t_barramin}.



\ifCLASSOPTIONcaptionsoff
  \newpage
\fi

\end{document}